\documentclass[twocolumn]{aastex631}

\usepackage{graphicx}	
\usepackage{amsmath}	
\usepackage{amssymb}	
\usepackage{color}
\usepackage{algorithm}
\usepackage{booktabs}
\usepackage[noend]{algpseudocode}
\usepackage{pifont}
\usepackage{soul}
\usepackage{float}
\usepackage{multibib}
\newcites{Appendix}{Appendix References}
\usepackage{natbib}
\usepackage[flushleft]{threeparttablex}
\usepackage{mathrsfs}
\usepackage{tikz}
\usepackage{esint}
\usepackage[draft]{todonotes}
\usepackage{lipsum}
\usepackage{longtable}
\usepackage{pifont}
\usepackage{savesym}
\savesymbol{tablenum}
\usepackage{siunitx}
\restoresymbol{SIX}{tablenum}
\usepackage{placeins}
\usepackage{bm}

\newcommand{\appropto}{\mathrel{\vcenter{
  \offinterlineskip\halign{\hfil$##$\cr
    \propto\cr\noalign{\kern2pt}\sim\cr\noalign{\kern-2pt}}}}}

\newcommand{\Exp}[2]{\left\langle{#1}\right\rangle_{#2}}

\renewcommand{\d}[1]{\ensuremath{\operatorname{d}\!{#1}}}

\DeclareMathOperator\D{\mathcal{D}}
\DeclareMathOperator\V{\mathcal{V}}
\newcommand{\M}{\mathcal{M}}

\DeclareMathOperator\s{\text{s}}
\DeclareMathOperator\f{\text{f}}

\DeclareMathAlphabet\mathbfcal{OMS}{cmsy}{b}{n}

\renewcommand{\Re}{{\rm{Re}}}

\renewcommand{\v}{\mathit{v}}
\renewcommand{\b}{\mathit{b}}

\renewcommand{\u}{\bm{u}}
\newcommand{\us}{\bm{u}_s}
\newcommand{\uc}{\bm{u}_c}
\renewcommand{\k}{\bm{k}}
\newcommand{\x}{\bm{x}}
\newcommand{\K}{\bm{K}}
\renewcommand{\P}{\mathcal{P}_{u}}
\newcommand{\bom}{\bm{\omega}}

\newcommand{\T}{\mathcal{T}}
\newcommand{\bnab}{\bm{\nabla}}

\newcommand{\baro}{\bnab\rho \times \bnab P/\rho^2}
\newcommand{\Pus}{\mathcal{P}_{\u_s}}
\newcommand{\PB}{\mathcal{P}_{\rm B}}
\newcommand{\PoB}{\mathcal{P}_{\omega \rm B}}
\newcommand{\Po}{\mathcal{P}_{\omega}}

\begin{document}

\correspondingauthor{\\
$^{\dagger}$James R. Beattie: \href{mailto:jbeattie@cita.utoronto.ca}{jbeattie@cita.utoronto.ca}}

\title{Supernovae drive large-scale, incompressible turbulence through small-scale instabilities}

\author[0000-0001-9199-7771]{James R. Beattie$^{\dagger}$}
\affiliation{Canadian Institute for Theoretical Astrophysics, University of Toronto, 60 St. George Street, Toronto, ON M5S 3H8, Canada}
\affiliation{Department of Astrophysical Sciences, Nassau Street, Princeton University, Princeton, NJ 08544, USA}

\begin{abstract}
    The sources of turbulence in our Galaxy may be diverse, but core-collapse supernovae (SNe) alone provide enough energy to sustain a steady-state galactic turbulence cascade at the observed velocity dispersion. By localizing and analyzing supernova remnants (SNRs) in high-resolution SN-driven galactic disk cut-out simulations from \citet{Beattie2025_SLK41}, I show that isolated SNRs source incompressible turbulence through baroclinic vorticity generation localized at the unstable contact discontinuity. Through the spherical harmonic power spectrum of the corrugations, I provide evidence that this process is seeded by surface instabilities and 2D turbulence on the shell, which corrugates and folds the interface, becoming the strongest source of baroclinicity in the simulations. I present an analytical relation for a baroclinicity-fed incompressible mode (co)spectrum, which matches that observed in the simulated SNRs, and reveals a $\propto k^{3/4}$ spectrum that drives the turbulence. I show that vortex stretching allows for modes to be shed from the contact discontinuity into the surrounding medium and derive a timescale criterion for this process, revealing that young SNR with radii close to the cooling radius are efficient at radiating turbulence. The unstable layer produces a spectrum of incompressible modes $\propto k^{-3/2}$ locally within the SNRs. Through the inverse cascade mechanism revealed in \citet{Beattie2025_SLK41}, this opens the possibility that the $k^{-3/2}$ spectrum, arising from corrugated folds in the unstable layer, imprints itself on kiloparsec scales, thereby connecting small-scale structure in the layer to the large-scale incompressible turbulence cascade.
\end{abstract}

\keywords{turbulence, hydrodynamics, ISM: kinematics and dynamics, galaxies: ISM, galaxies: structure}

\section{Introduction} \label{sec:intro}
    The vast length scales within a galaxy, combined with the very low viscosities of warm and cold plasma in the interstellar medium \citep{Ferriere2020_reynolds_numbers_for_ism}, mean that even moderate velocity perturbations can trigger a turbulence cascade across all inviscid modes. Turbulence creates self-similar velocity dispersion relations that couple the root-mean-squared velocities across vast scales \citep{Armstrong1995_power_law,Federrath2021,Colman2022_large_scale_driving_in_ISM,Beattie2025_nature_astro,Connor2025_cascading_from_the_winds}. It dynamically mixes metals and dust, and fragments the mass density \citep{Hopkins2013_fragmentation,Federrath2016_filaments,Beuther2015,Krumholz2018_metallicity_SF,Kolborg2022_metal_mixing_1}. Further, turbulence fuels the magnetic fluctuation dynamo, creating intensely folded magnetic field geometries \citep{Schekochihin2004_dynamo,Federrath2014_supersonic_dynamo,Seta2022_multiphase_dynamo,Kriel2022_kinematic_dynamo_scales,Kriel2025_SSD,Gent2021_supernova_turbulence_and_dynamo}, which in turn couples to cosmic-ray scattering, diffusion, and transport in the disk and winds \citep{Sampson2023_SCR_diffusion,Hopkins2021_cr_transport_models,Kempski2023_b_field_reversals,Ruszkowski2023_CRs_in_galaxies_review,Sampson2025_CRMHD_heating,Lubke2025_CR_transport}. Indeed, any light that travels to us has to pass through this turbulent medium, and hence signatures of the medium imprint themselves on the light, whether it be structures generated by the turbulence or the eddies themselves  \citep{Armstrong1995_power_law,Jow2024cuspcuspsuniversalmodel,Ocker_2022_radio_scattering,kempski2024unifiedmodelcosmicray}. It is difficult to truly encapsulate the vast array of roles that turbulence plays within a galaxy.    
    
    Core-collapse supernovae (SNe) have more than enough energy to be one of the primary drivers of a galactic turbulence cascade \citep{MacLow2004,Hill2012_SNe_driven_turb,Beattie2025_SLK41,Connor2025_cascading_from_the_winds}. Simulations have demonstrated that SNe create and sustain steady-state turbulence \citep{Hennebelle2014_SN_driven_turb,Padoan2016_supernova_driving,Bacchini2020_supernova_drives_turb,Gent2021_supernova_turbulence_and_dynamo,Gent2022_multiphase_dynamo,Beattie2025_SLK41,Connor2025_cascading_from_the_winds}. However, SN blast waves alone are insufficient to excite even weak acoustic turbulence between themselves, let alone Kolmogorov-style turbulence \citep{Mee2006_blastwave_turbulence}. The prevailing theory is that turbulence is a byproduct, excited through corrugated or curved shock waves in blast-wave interactions, in collisions between hot superbubbles, and in interactions with inhomogeneities in the medium \citep{Elmegreen2004,Balsara2004_SNe_driven_turb,Kapyla2018_vorticity_helicity_ISM}, which I challenge in this Letter.
    
    By analyzing a population of supernova remnants (SNRs) extracted as they expand through a homogeneous ISM from global SN-driven turbulence simulations, in this Letter I show that interactions between multiple SNe and ISM inhomogeneities are not required to generate turbulence. Instead, SNRs generate turbulence via baroclinicity from the unstable thin shell at the interface between warm and hot plasma. The layer becomes fractal and folded, with a $\PB(k) \propto k^{3/2}$ baroclinic spectrum (similar to a kinematic dynamo spectrum), due to what I argue is seeded by interface instabilities \citep{Vishniac1983_overstability,MacLow1993_overstability,McLeod2013_simulations_of_the_nonlinearVish,Badjin2021_SNR_instabilities}, and then further corrugated by 2D surface turbulence, creating power law spectra in the spherical harmonics of the layer. At the sites of the corrugated layers in the SNRs, an incompressible velocity spectrum develops, following $\Pus(k) \propto k^{-3/2}$, which is purely sourced by the baroclinicity of the layer. This is the same spectrum observed in global SN-driven turbulence simulations that extend from the disk into the galactic winds \citep{Padoan2016_supernova_driving,Fielding2018_winds,Beattie2025_SLK41,Connor2025_cascading_from_the_winds}. Recently, \citet{Beattie2025_SLK41} showed that through incompressible–compressible–incompressible triad velocity–mode interactions, SN-driven turbulence generates an inverse cascade from the cooling radius of an expanding SNR, where the unstable interface forms, into the galactic winds. In the present study, I show that the same triad is responsible for shedding incompressible turbulence from the thin shell and ejecting it into the surrounding medium. Hence, I suggest that the large-scale $\Pus(k) \propto k^{-3/2}$ spectrum originates from instabilities and turbulence on the unstable shell, which is then transported to the outer scales of the galaxy via \citet{Beattie2025_SLK41}'s inverse cascade.

    This Letter is organized as follows. In \autoref{sec:sims} I describe the numerical simulations, the method used to extract SNRs from the simulations of \citet{Beattie2025_SLK41}, and the definitions of the Fourier power spectra used in my analysis. In \autoref{sec:generation_of_vorticity} I provide an analytical prediction for the scaling between the vorticity–baroclinic cospectrum (the driving spectrum) and the incompressible mode spectrum, assuming that baroclinicity directly sources vorticity at the SNR scale. I show that this relation holds in the SNR models and that the incompressible velocity spectrum follows $k^{-3/2}$ at the SNR scale, and the spectrum that drives the turbulence, $k^{3/4}$. In \autoref{sec:evidence_for_instability} I provide an analytical model for how the baroclinic-generated modes are shed from the surface of the thin shell, providing a criterion based on the 2D confinement timescale and the three-dimensionalization timescale. I show SNRs that are close to their cooling age are the most efficient at generating and shedding turbulence into the surrounding medium. In this section, I also perform a full characterization of the spherical harmonics of the radial fluctuations of the layer, which indicates that the thin shell supports 2D turbulence. Finally, in \autoref{sec:summary_and_conclusions} I summarize the key results and discuss the broader implications of my results for SN-driven turbulence in galaxies.

\begin{figure*}
    \centering
    \includegraphics[width=\linewidth]{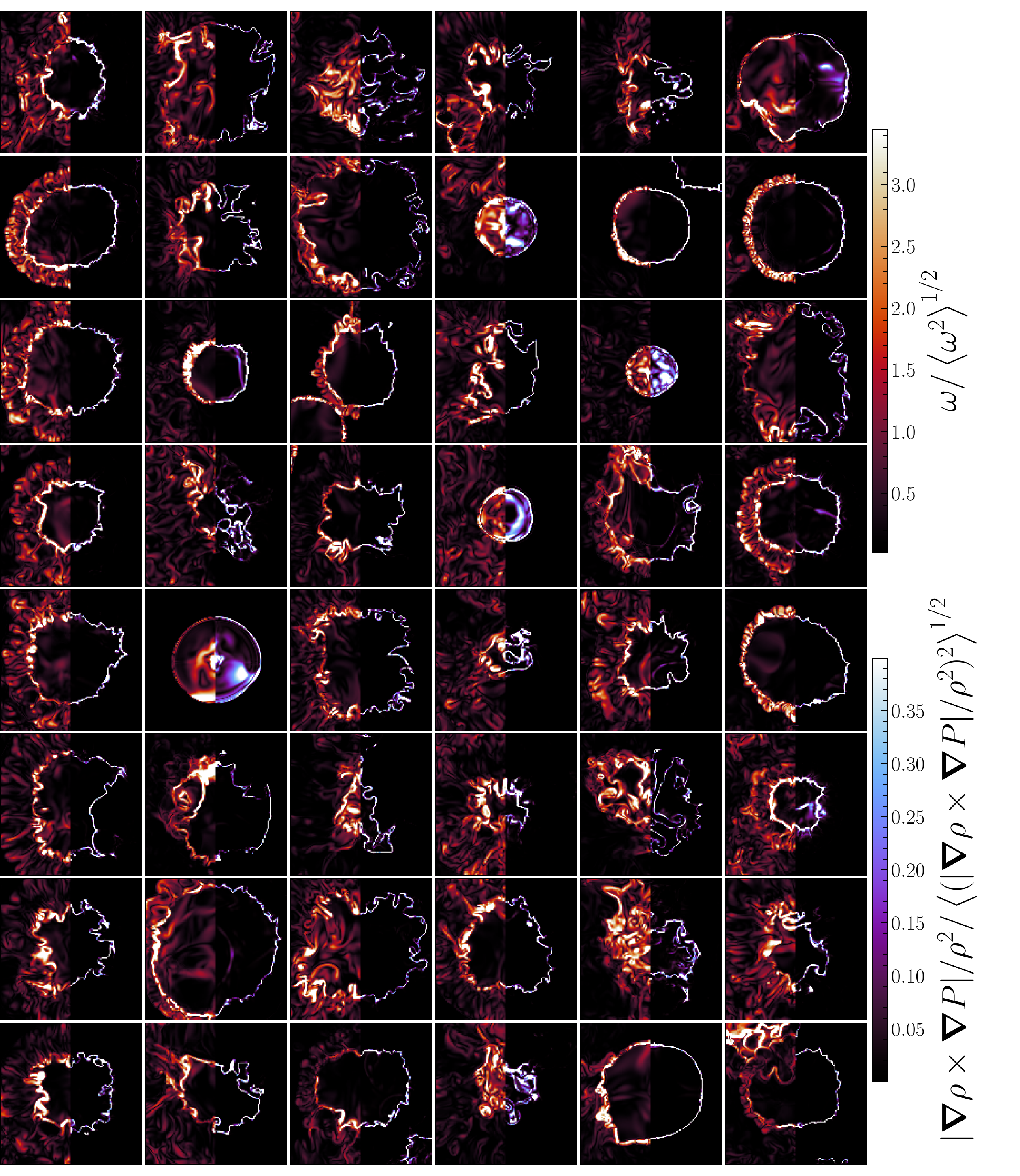}
    \caption{Two-dimensional slices of the root-mean-square normalized vorticity, $\omega/\langle \omega^2 \rangle^{1/2}$ ($\omega = |\bnab \times \u|$; left in each panel), and baroclinicity ($\baro$, right in each panel), for 48 randomly selected supernova remnants (SNRs) extracted from SN-driven turbulence simulations in \citet{Beattie2025_SLK41}. The most intense $\omega$ regions, which are signatures of incompressible turbulence, correspond to the strongest $\baro$ structures. The $\baro$ structures closely trace the corrugated layer between the warm and hot plasma. \citet{Beattie2025_SLK41} finds that this layer generates $\bom$ values three orders of magnitude larger than any other vorticity source in an SN-driven galactic disk environment. In this study, I show that these layers drive incompressible turbulence with a velocity spectrum $\propto k^{-3/2}$, the same spectrum as what is found in global SN-driven turbulence simulations of a galactic disk \citep{Connor2025_cascading_from_the_winds}. An animation of this panel with a slider for the vertical bar between the vorticity and baroclinicity can be found \href{https://astrojames.github.io/movies/}{here}.}
    \label{fig:vort_baro_slices}
\end{figure*}

\section{Numerical Methods}\label{sec:sims}
\subsection{Gravitohydrodynamical stratified, multiphase, SN-driven turbulence model}
    For a detailed description of the numerical setup analyzed in this study, see \citet{Beattie2025_SLK41}; see also \citet{Kolborg2022_metal_mixing_1,Kolborg2023_metal_mixing_2} and \citet{Martizzi2016}. Briefly, I use the \textsc{ramses} code \citep{Teyssier2002_ramses} to simulate ideal (no explicit viscosity), stratified, gravitohydrodynamical SN-driven turbulence, with a time-dependent cooling network to model the large-scale, volume-filling phases of the ISM (WNM, WIM, HIM). The setup excludes self-gravity, magnetic fields, cosmic rays, and large-scale galactic shear (see the final section of \citealt{Beattie2025_SLK41} for a comprehensive discussion of model limitations, including the cooling function that neglects the cold phase plasma). This makes the simulation an extremely controlled numerical experiment, designed to isolate the impact of SN explosions on stratified, multiphase turbulence. The domain is periodic in the plane perpendicular to the static gravitational potential gradient, $\partial_z \phi$, with outflow boundaries that allow galactic winds to escape, though these remain minimal \citep{Martizzi2016}. The SN explosion rate is set to $\gamma_{\rm{SNe}}/10^{-4}\,\rm{yrs} = 0.1$, and the parameterization of $\phi$ and the total mass are chosen such that the galaxy model is roughly a Milky Way analogue. A study of different SN prescriptions, across different galactic models, can be found in \citet{Connor2025_cascading_from_the_winds}. The simulation uses a domain discretization of $1\,\!024^3$ on $L^3 = 1\,(\rm{kpc})^3$, which resolves scales from $1\,\rm{kpc}$ down to $\sim 10\,\rm{pc}$ before numerical diffusion truncates the dynamics \citep{Malvadi2023_numerical_diss,Grehan2025_num_diffusion,Beattie2025_bulk_viscosity}. Throughout the study, I use $\u$, $\rho$, $P$, and $T$ to denote the fluid velocity, mass density, pressure, and temperature, respectively. 

\subsection{Localizing SNR and building a catalog}
    I focus on studying the local Fourier spectral properties of SNR, as opposed to the global treatment in \citet{Beattie2025_SLK41} and other SN-driven turbulence studies. This requires grouping and identifying the SNR on the simulation grid. In this study I do not track the clusters in time, which simplifies the algorithm. I summarize the basic steps below.
    
    SNe drive localized hot bubbles of plasma with $T \gg 10^6\,\rm{K}$. Hence, to identify the SNR I first create a binary mask of the temperature field, $T$, such that $\log T / \left\langle T^2\right\rangle^{1/2} > \tau_{\rm temp}$. Because hot shells of plasma beyond the cooling radius of the SNR are also captured by this mask, I apply a binary opening filter, commonly used in computer vision \citep{Dawson-Howe2014}, to ensure that any hot filaments are disconnected from the bulk of the SNR. After processing the mask in this way, I apply a friend-of-friends (FoF) algorithm\footnote{For this simulation, I use $\tau_{\rm temp} = -1.5$. The full implementation can be found  \href{https://github.com/AstroJames/PLASMAtools/blob/main/funcs/clustering/examples/main_grid_demo.py}{here}, which includes additional capabilities beyond what are discussed in this section.}, similar to \citet{Robertson2018}. I use the FoF to identify $n$ distinct, spatially grouped hot regions across several realizations during the first few turbulent turnover times in the simulation (see \autoref{app:clusters} and \autoref{fig:clusters} for a visualisation of the identified SNR in the global simulation). This ensures that I am probing SNRs that are: (1) expanding in background medium that is mostly homogeneous (see the left panels in Figure~4 and Figure~11 in \citealt{Beattie2025_SLK41}), and (2) that there has not been many interactions between different forward shocks from the SN explosions. This allows me to isolate the effects of how the unstable layer in the SNR drives incompressible turbulence, which is the main goal of this study. Finally, I discard FoF regions with volume $\V_i < \tau_{\rm vol}$, thereby removing low–volume-filling shells from the forward shocks which I do not want to extract\footnote{I set $\tau_{\rm vol}$ to 500 cells, which removes a lot of the spurious regions far away from any SNR that I do not want to include in my SNR catalog.}. The exact thresholds require tuning for simulations at different resolutions but overall the algorithm is quite simple, reasonably robust and scales adequately such that it can easily be used on $1,\!000^3$ grids.
    
    This approach is successful in identifying SNRs but not sufficient for full automation. I therefore review the final output and select $\approx 100$ SNRs ($n = 87$) by extracting $128^3$ grid-cell ($\ell_0 = 125\,\rm{pc}$) regions around each candidate, centered on the geometric mean of the FoF region, after visually inspecting 2D temperature slices. Full automation would be possible with additional effort, but this procedure is sufficient for constructing a statistical sample for the purposes of this study. These $\approx 100$ SNRs form the dataset used in the remainder of this work. No statistics in the main text are reported from the global simulations -- everything is local to the SNRs.
    
    I show a random sample of 48 two-dimensional slices through the three-dimensional SNRs. I plot the vorticity, $\omega = |\bnab \times \u|$, and baroclinicity, $|\baro|$, with $\omega$ shown in red (left) and $|\baro|$ in blue (right). $|\baro|$ traces the unstable contact discontinuity between the two plasma temperatures very closely, allowing it to be used directly to analyze the structure of the layer. In contrast, $\omega$ is strong both at the layer coinciding with the largest $|\baro|$, and (less so) outside the SNR. Because I use early snapshots from the global simulations to isolate the unstable layer in individual SNRs, most of the $\omega$ is generated and sourced by the layer itself (see in \autoref{app:baroclinic_source} that this is true for the steady state too, and the shock-shock interactions and curved shocks end up playing a negligible role in driving the turbulence, in general). I now turn to analyzing the Fourier-mode power spectra of the layer in detail.

\begin{figure*}
    \centering
    \includegraphics[width=\linewidth]{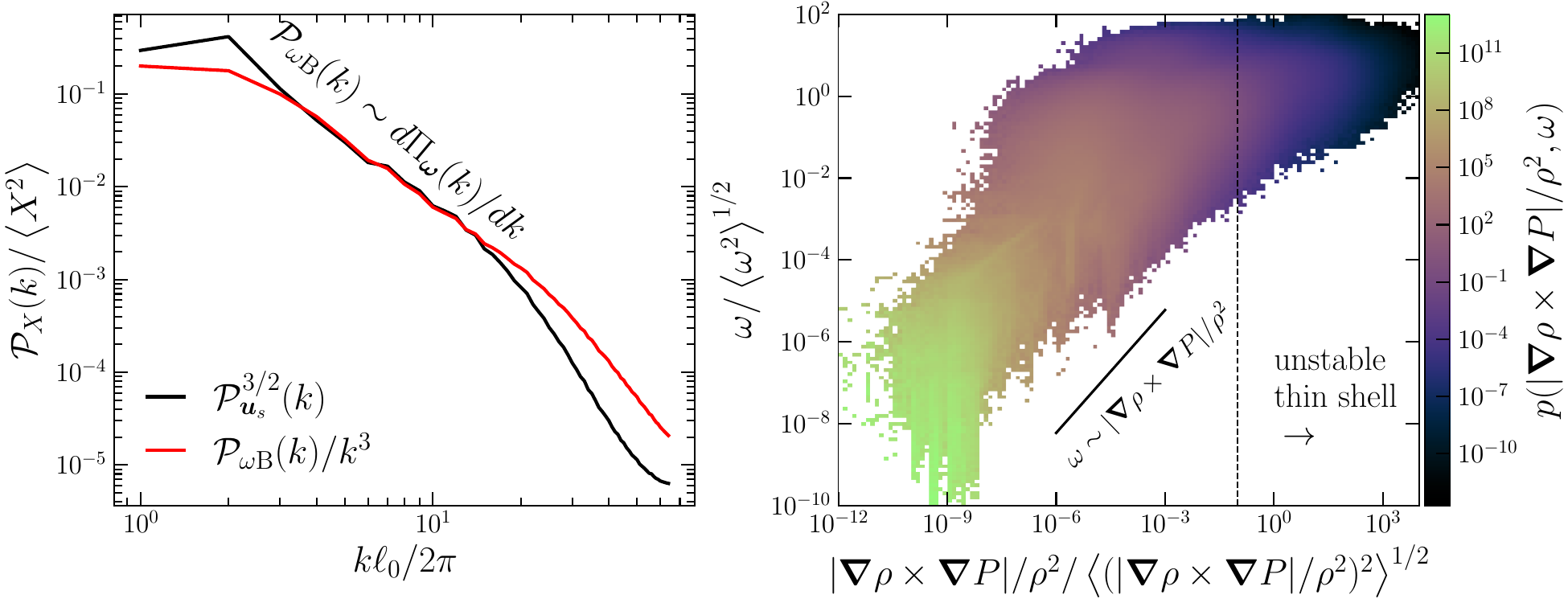}
    \caption{\textbf{Left:} The vorticity–baroclinic power (co)spectrum, $\PoB(k)$ (\autoref{eq:ps_ob}), and the incompressible velocity-mode power spectrum, $\Pus(k)$ (\autoref{eq:ps_sol}), averaged over $\approx 100$ SNRs and normalized to test \autoref{eq:baro_sol_relation}. Because the two transformed spectra scale with each other almost perfectly, the enstrophy sourced by baroclinicity, $\PoB$, fully accounts for the enstrophy flux entering the cascade, $d\Pi_{\bm{\omega}}/dk$. This demonstrates that the low-volume, fractal layer, which dominates the baroclinicity (see \autoref{app:baroclinic_source} for a more detailed calculation showing that the unstable layer alone provides between all and 70\% of the baroclinicity in the global simulations), drives the incompressible turbulence across a broad band of modes. \textbf{Right:} The two-dimensional volume-weighted probability density function (PDF) of vorticity and baroclinicity, showing a strong positive correlation, $\omega \sim |\baro|$, that peaks and flattens at the $\baro$ values concentrated within the fractal layer (see \autoref{fig:vort_baro_slices}). This shows that volume-poor layers can dominate the baroclinic production.}
    \label{fig:baro_enstrophy_spectra}
\end{figure*}

\begin{figure}
    \centering
    \includegraphics[width=\linewidth]{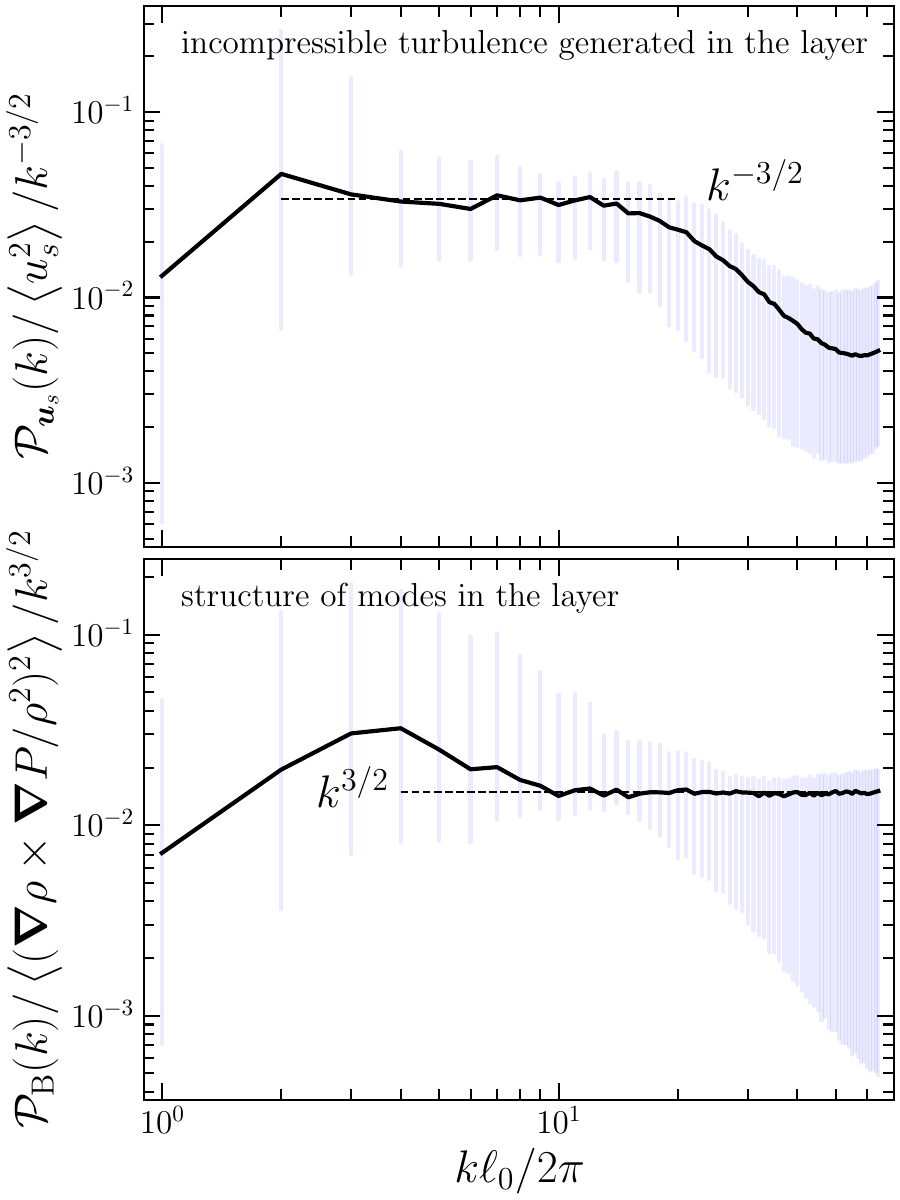}
    \caption{\textbf{Top:} The incompressible velocity-mode spectrum, $\Pus(k)$, averaged (black) over the ensemble of localized SNRs and normalized by both the integral of the spectrum and a $k^{-3/2}$ compensation. The ensemble $1\sigma$ is shown in transparent blue. The wavenumber is normalized to $\ell_0 = 125\,\rm{pc}$, the domain size of the region extracted around each SNR. The plot shows the development of a self-similar range of modes that already exhibit a $\Pus(k) \propto k^{-3/2}$ spectrum on the scales of individual SNRs. This is the same spectrum found in global disk cut-out simulations of SN-driven turbulence \citep{Padoan2016_supernova_driving,Beattie2025_SLK41,Connor2025_cascading_from_the_winds}.
    \textbf{Bottom:} The pure baroclinic power spectrum, $\PB(k)$ (\autoref{eq:ps_b}), normalized in the same way as the top panel but with a $k^{3/2}$ compensation. This spectrum probes the organization of Fourier modes in the fractal layer between the warm and hot plasma in the SNR. It exhibits a $\PB(k) \propto k^{3/2}$ power law, peaking at high-$k$, reflecting the signature of a highly folded layer structure with strong gradients on very small scales \citep{Schekochihin2004_dynamo}.}
    \label{fig:spectra}
\end{figure}
    
\section{Baroclinicity and incompressible mode correspondence}\label{sec:generation_of_vorticity}
    The principal goal of this study is to address the correspondence and origin of incompressible turbulence in an SN-driven system. \citet{Beattie2025_SLK41} argued that this turbulence does not arise from blast-wave interactions, as conjectured by \citet{Balsara2004_SNe_driven_turb}, but instead from baroclinicity, $\baro$, generated in the cooling layer of the SNR. In \autoref{app:baroclinic_source}, I show that even though the layer is volume-poor, it makes up almost all of the $\baro$ production at the onset of the turbulence, and $\gtrsim 70\%$ in steady-state. This occurs through misaligned radial $\bnab P$, from the explosion, combined with instabilities in the layer that produce density corrugations and misalignment with $\bnab \rho$ \citep{Badjin2021_SNR_instabilities}. I qualitatively demonstrate this correspondence for the 48 SNRs in \autoref{fig:vort_baro_slices} (also shown with animations here \href{https://astrojames.github.io/movies/}{https://astrojames.github.io/movies/}).  

    \subsection{Driving the turbulence with baroclinicity generated by the unstable thin shell}
    
    In this study I analyze several isotropically integrated power spectra over the extracted SNR regions. The first is the incompressible (solenoidal; $\bnab\cdot\u_s = 0$) velocity power spectrum, 
    \begin{align}\label{eq:ps_sol}
        \Pus(k) = \int k^2 \, \d{\Omega_k} \, \u_s(\k)\cdot\u^\dagger_s(\k),
    \end{align}
    where $\u_s(\k)$ is the Fourier-transformed incompressible velocity field, constructed via the Helmholtz decomposition described in \citet{Beattie2025_SLK41,Connor2025_cascading_from_the_winds}, $\u^\dagger_s(\k)$ is its complex conjugate, and $\d{\Omega_k}$ is the solid angle at fixed $k = |\k|$. The second is the mixed cospectrum between baroclinicity and vorticity,
    \begin{align}\label{eq:ps_ob}
        \PoB(k) = \int k^2 \d{\Omega}_k \, |\Re\left\{\bom(\k)\cdot (\baro)^\dagger(\k)\right\}|,
    \end{align}
    where $\Re\{\hdots\}$ denotes the real part. This form follows directly from the Fourier transformed enstrophy ($\omega^2$) evolution equation,
    \begin{align}
    \frac{1}{2}\frac{\partial \bom(\k)\cdot\bom^{\dagger}(\k)}{\partial t} =& -\frac{d\Pi_{\omega}(\bm{k})}{d\k} \nonumber \\ 
    +& \Re\left\{\bom(\k)\cdot(\baro)^{\dagger}(\k)\right\} \nonumber \\
    -& \mathcal{D}(\k), 
    \end{align}
    $d\Pi_{\bom}(\k)/d\k$ is the enstrophy transfer function for the cascade terms, and $\mathcal{D}(\k)$ for the diffusion terms. Hence \autoref{eq:ps_ob} directly probes the spectrum of transfer from $\baro$ into $\bom$, which links $\baro$ to the generation of the incompressible turbulence since $\omega \sim ku_s$ (\autoref{app:spectrum_vort}). Finally, to probe the structure of the unstable layer itself, I analyze the baroclinic spectrum, 
    \begin{align}\label{eq:ps_b}
        \PB(k) &= \nonumber \\
        &\int k^2 \d{\Omega}_k \, (\baro)(\k)\cdot (\baro)^\dagger(\k),
    \end{align}
    noting that $\baro$ closely traces the layer structure, as shown in \autoref{fig:vort_baro_slices} and throughout this study (see also \citealt{Beattie2025_SLK41}).  

    In statistical steady state, the isotropically integrated, Fourier-transformed enstrophy evolution equation ($\partial_t \Exp{\int k^2 \d{\Omega}_k\, \bom(\k)\cdot\bom^{\dagger}(\k)}{} = 0$) is
    \begin{align}
         0 = -\Exp{\frac{d\,\Pi_{\bom}(k)}{dk}}{} + \Exp{\PoB(k)}{} - \Exp{\mathcal{D}(k)}{},
    \end{align}
    where $\Exp{\hdots}{}$ is taken over the ensemble of SNRs\footnote{All the statistics I report are averaged over a full ensemble of SNR, hence I may invoke stationarity to set the time-derivatives to zero.}. If all the flux sourced from $\baro$ is proportional to the flux feeding the incompressible velocity cascade, then
    \begin{align}
        \Exp{\PoB(k)}{} &= \Exp{\frac{d\,\Pi_{\bom}(k)}{dk}}{} \nonumber \\
        &\sim \frac{1}{t_{\rm nl}}\Exp{\int \d{\Omega}_k \, k^2 \, \bom(\k)\cdot\bom^\dagger(\k)}{} 
        \sim \frac{\Exp{k^2 \mathcal{P}_{\u_s}(k)}{}}{t_{\rm nl}}, 
    \end{align}
    where $t_{\rm nl} \sim (k u_k)^{-1}$ is the nonlinear timescale of the turbulence, with $u_k^2 \sim \Pus(k)dk$. This gives $t_{\rm nl} \sim (k \sqrt{ \mathcal{P}_{\u_s} dk})^{-1}$, and hence
    \begin{align}\label{eq:baro_sol_relation}
          \PoB(k) &\sim k^3 \Pus^{3/2}(k),
    \end{align}
    or equivalently, $\Pus(k) \sim  \PoB^{2/3}(k)\,k^{-2}$. This relation between the spectra holds only if the entire enstrophy cascade, $d\Pi_{\bom}(k)/dk$, e.g., the incompressible turbulence, is sourced by $\baro$. Qualitatively from \autoref{fig:vort_baro_slices} and quantitatively from \autoref{app:baroclinic_source}, we understand this, to leading order, comes directly from the thin, corrugated layer between the hot and warm plasma in and around the SNR (the unstable contact discontinuity).

    I plot $\PoB(k)/k^{3}$ (red) and $\Pus^{3/2}(k)$ (black) in the left panel of \autoref{fig:baro_enstrophy_spectra}, averaged over all SNR and normalized by the integral of each spectrum. The curves trace each other almost perfectly, confirming \autoref{eq:baro_sol_relation} and indicating that the $\baro$ generated in the SNR cooling layer directly fuels the $\omega$ cascade, thereby sourcing incompressible (solenoidal) velocity turbulence\footnote{Note that this means there is no driving scale for this type of turbulence, but instead an entire driving spectrum that peaks at high-$k$ $\PoB(k) \propto k^{3/4}$.}. I further show the correlation between $\omega$ and $|\baro|$ in the right panel of \autoref{fig:baro_enstrophy_spectra}, where I plot the two-dimensional probability density function (PDF) of $\omega$ and $|\baro|$. The PDF reveals a linear correlation at low $\omega$ and low $|\baro|$ (as annotated on the plot), which implies extremely rapid, kinematic-like growth of $\omega$ from $\baro$, since $\partial_t \bom \sim \baro \sim \bom$. Moreover, the most extreme values of $\omega$ occur within the $|\baro|$ range associated with the thin layer, as shown in \autoref{fig:vort_baro_slices}. This is not unlike other regimes of turbulence where intermittent structures can occupy very little volume but contribute in large-ways to the dissipation \citep{Falgarone1995,Falgarone2009,Zhadankin2013_current_sheet_statistics}. However, in this case it is intermittent structures contributing to driving (source term) rather than dissipating (sink term) the turbulence. 

\begin{figure*}
    \centering
    \includegraphics[width=\linewidth]{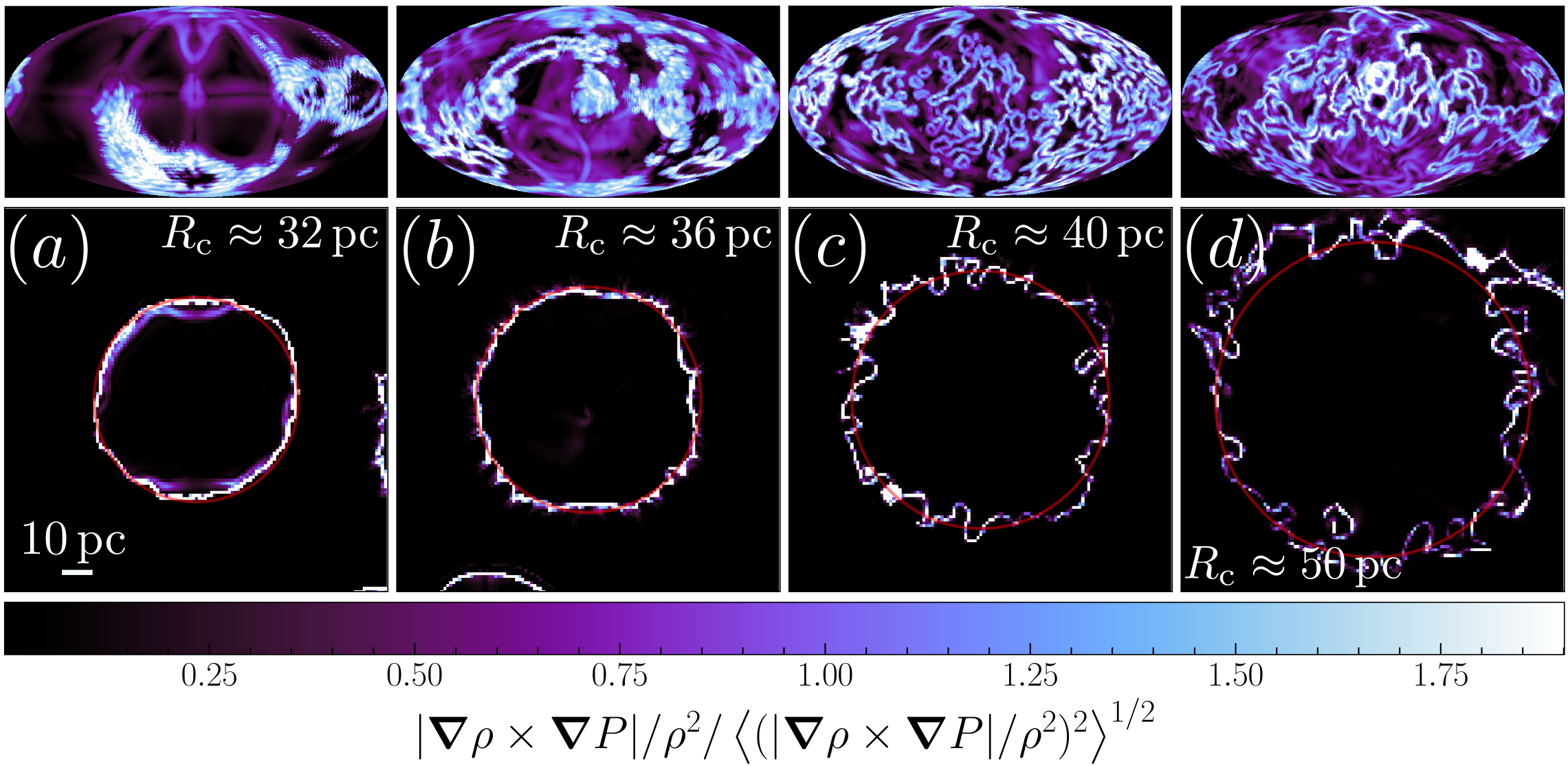}
    \caption{
    Evolution of the unstable thin shell traced by $\baro$ in the SNR. Each column from $(a)$ to $(d)$ shows the SNR at an increasing radius, $R_{\rm c}$ ($\sim$ age; indicated in red in the bottom row), all taken from the same early-time realization of the global simulation to ensure only internal instabilities impact the layer. As the shell expands, high-$k$ modes grow in the layer. By the time it reaches the full $125\,\rm pc$ domain size, the shell has become highly fractal, with deep, folded corrugations, without any background inhomogeneities to enhance or modify the unstable modes. \textbf{Top row:} the $\baro$ ``sky'' for an observer at the center of the SNR, sliced at the radius, $R_{\rm c}$, highlighted in red in the bottom row, qualitatively showing that high-order spherical harmonics become increasingly excited as the SNR expands. \textbf{Bottom row:} two-dimensional slices of $\baro$ (the same as in \autoref{fig:vort_baro_slices}), showing the evolving, expanding corrugated layer.
}
    \label{fig:SNR_timeline}
\end{figure*}

    \subsection{The incompressible mode spectrum generated at the SNR scale, by the unstable layer}\label{sec:the_incomp_spectrum}

    I show the incompressible mode spectrum, $\Pus(k)$, generated by the unstable layer in the top panel of \autoref{fig:spectra}, averaged over the full SNR ensemble, with $1\sigma$ over the ensemble shown in blue. Because $\ell_0 = 125\,\rm{pc}$ is discretized with $128^3$ cells, we can expect numerical diffusion effects for $k\ell_0/2\pi \gtrsim 10$, which is evident in the exponential truncation of the spectrum at high $k$. Before this truncation, the spectrum follows a $\Pus(k) \propto k^{-3/2}$ power law, which I confirm by compensating $\Pus(k)$ by $k^{-3/2}$ and showing that the spectrum flattens completely. A $\Pus(k) \propto k^{-3/2}$ spectrum is particularly interesting because it appears ubiquitously across a number of different regimes: in magnetohydrodynamics \citep{Iroshnikov_1965_IK_turb,Kraichnan1965_IKturb,Boldyrev2006,Beattie2025_nature_astro,Beattie2025_sda}, in hydrodynamical and magnetohydrodynamical acoustic and fast-wave (weak) turbulence \citep{Galtier2023_fast_wave_turb,Kochurin2024_weak_versus_strong_turb}, and in global SN-driven simulations with and without magnetic fields \citep{Padoan2016_supernova_driving,Beattie2025_SLK41,Connor2025_cascading_from_the_winds}. Indeed, \citet{Connor2025_cascading_from_the_winds} presented evidence for a velocity spectrum $\mathcal{P}_{u}(k) \propto k^{-3/2}$ that smoothly connects turbulence in the galactic winds to that in the disk. The novelty of my result is that a $\Pus(k) \propto k^{-3/2}$ spectrum emerges on the scale of individual SNRs, and, as shown in \autoref{fig:baro_enstrophy_spectra} and \autoref{app:baroclinic_source}, is entirely sourced by $\baro$ in the unstable thin shell between the hot and warm plasma. This suggests a link between the global SN-driven turbulence and the unstable modes that cause the folds and corrugations of the layer, which inevitably control the $\baro$ turbulence driving. 

\section{Mode structure of the unstable shell}\label{sec:evidence_for_instability}
    \subsection{Geometry of the generated modes: are the modes stuck on the thin shell?}\label{ssec:stuck_on_shell}

    In this section I work in spherical coordinates of the thin shell, $(\hat{\bm{r}},\hat{\bm{\theta}},\hat{\bm{\varphi}})$, where $r=0$ is at the center of the SNR and $\theta$ and $\varphi$ are the standard azimuthal and polar angles. Because the corrugations are in $\hat{\bm{r}}$ it is further useful to define the radial displacement vector, $\xi_r$, where corrugations are $\hat{\bm{\theta}},\hat{\bm{\varphi}}$ fluctuations in $\xi_r$,
    \begin{align}\label{eq:displacement_vector}
         \delta \xi_r(\theta,\varphi) = \xi_r(\theta,\varphi) - \xi_{r,\rm iso},
    \end{align}
    where $R_{\rm c} \equiv \xi_{r,\rm iso} = \Exp{\xi_r(\theta,\varphi)}{\theta,\varphi}$ is a mean radius over $\theta$ and $\varphi$. I define this as the radius of the contact discontinuity, $R_{\rm c}$.
    
    Let me consider a SNR with dominant radial pressure $\bnab P \approx \partial_r P \hat{\bm{r}}$ and corrugations $\bnab \rho \approx \bnab_\perp \delta \xi_r$, where $\bnab_\perp$ is the gradient operator perpendicular to $\hat{\bm{r}}$. Then $\hat{\bom} \approx \bnab_\perp \delta \xi_r \times \hat{\bm{r}}$ produces $\hat{\bom} \,\|\, \bnab_\perp \delta \xi_r \times \hat{\bm{r}}$ in the tangent plane on the surface of the thin shell, orthogonal to $\bnab_\perp$. The direction of the turbulent modes it develops, $\us$, is then given by $\us(\k) \propto k\bm{\hat{r}} \times \bom(\k)$, and hence $\hat{\bm{r}}$, $\hat{\bom}$ and $\hat{\us}$ form an orthonormal basis on the surface, where $\hat{\bom}$ and $\hat{\us}$ span the tangent plane. This naturally facilitates thin layers, $\delta \xi_r/R_{\rm c} \ll 1$ of $\us$ modes that are confined along the shell. A natural question is then how do the modes confined on the surface ever interact with the surrounding medium?
    
    As \citet{Beattie2025_SLK41} showed, the $\us$ modes generated by $\baro$ at the layer interact with $\uc \approx u_c\hat{\bm{r}}$ from the SNR explosion. They stretch $\us$ modes through $\us\otimes\uc:\bnab\otimes\us$, which facilitates the inverse cascade measure in \citet{Beattie2025_SLK41}. They also contribute directly to three-dimensionalizing the flow, which I show in the following. The radial vorticity couples to the tangent plane velocities and vorticity via vortex stretching, 
    \begin{align}
        \frac{\partial \omega_r}{\partial t} = \frac{\omega_\theta}{r} \frac{\partial u_r}{\partial \theta} + \frac{\omega_\varphi}{r \sin\theta}\frac{\partial u_r}{\partial\varphi} - \frac{1}{r}\left(\omega_\theta u_\theta + \omega_\varphi u_\varphi\right),
    \end{align}
    hence the $\bom_{\theta,\varphi}$ generated on the surface can always be ejected into surrounding layer via coupling to the $u_r\hat{\bm{r}} \approx u_c\hat{\bm{r}}$. Taking a local patch on the surface, $\omega_{\theta,\varphi} \approx \omega_t$, $\partial_\varphi \sim k_\perp R_{\rm c}\sin\theta$ and $\partial_\theta \sim k_\perp R_{\rm c}$, then
    \begin{align}\label{eq:vortex_stretching}
        \frac{\partial \omega_r}{\partial t} \sim \omega_t k_\perp u_r - \frac{\omega_t u_t}{R_{\rm c}}.
    \end{align}
    For small corrugations, such that $k_\perp \gg 1/R_{\rm c}$, $\partial_t \omega_r \sim \omega_t k_\perp u_r$. This is very useful. First it gives a timescale for how fast the surface modes become 3D, $t_{\rm 3D} \sim \omega_t / (\partial_t \omega_r) \sim 1/ (k_\perp u_r)$, and second it shows that high-$k_\perp$ modes are the most efficient at shedding $\bom$ into the surrounding medium. Clearly, once 
    \begin{align}\label{eq:confinment_times}
     t_{\rm 3D} \sim t_{\rm nl} \sim \frac{1}{k_\perp u_t}, \quad\text{i.e.,}\quad \frac{t_{\rm nl}}{t_{\rm 3D}} \sim \frac{u_r}{u_t} \sim 1,   
    \end{align}
    where $t_{\rm nl}$ is nonlinear timescale of the $k_\perp$ surface modes, the 2D modes can be ejected into the surrounding medium on comparable timescales to being cascaded in 2D on the shell. I have assumed that the Reynolds number is sufficiently large on the shell and that the surface modes will dissipate via a cascade, given the chance. Hence the $t_{\rm nl}/t_{\rm 3D}$ provides a criterion for confinement and 2D interactions in the shell and ejection or shedding into the surrounding medium.
    
    In \autoref{fig:confinement_time} I plot \autoref{eq:confinment_times} for SNRs grouped by different $R_{\rm c}$ (see \autoref{app:confinement_time} for methodologies). Critically, I show that for $R_{\rm c} \lesssim 50\,\rm{pc}$, they are in a vortex shedding state with $t_{\rm nl}/t_{\rm 3D}\approx 10 > 1$. In contrast, $R_{\rm c} \gtrsim  60\,\rm{pc}$ SNRs, are in a confined state $t_{\rm nl}/t_{\rm 3D} < 1$. Hence, even though $\baro$ generates surface modes that are in the tangent plane of the shell, with large $u_r$ and short $t_{\rm 3D}$, the incompressible turbulence is able to radiate into the ISM through young\footnote{By young I mean $t_{\rm age} \approx t_{\rm cool} \approx 10^4-10^5\,\rm{yrs}$, and $R_{\rm cool} \approx 10-30\,\rm{pc}$, for typical ISM conditions \citep{Beattie2025_SLK41}.} SNRs.   

\begin{figure}
    \centering
    \includegraphics[width=\linewidth]{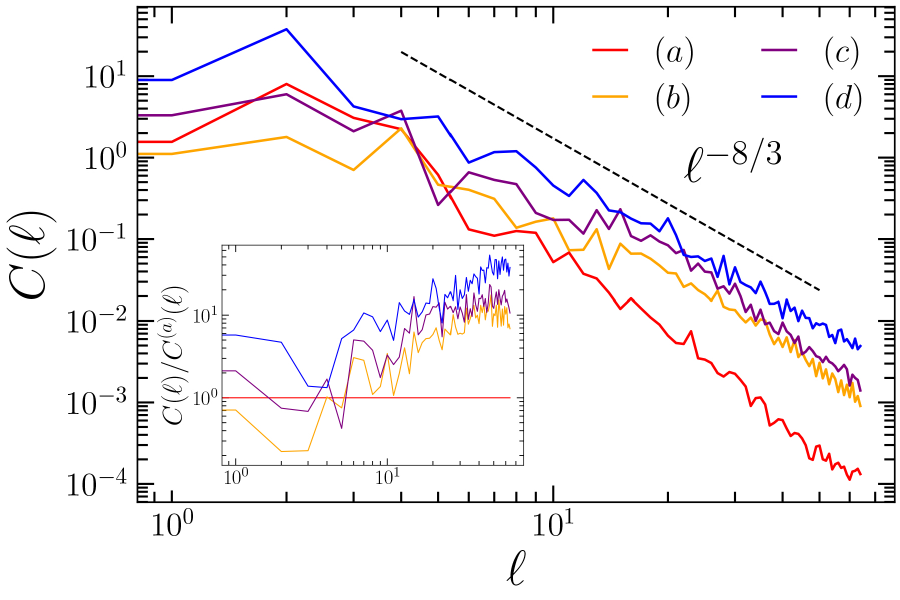}
    \caption{\textbf{Main:} the spherical harmonic power spectrum, $C(\ell)$, \autoref{eq:spherical_ps}, for the fluctuations in the radial displacement vector, $\delta \xi_r$, \autoref{eq:displacement_vector}. $\delta \xi_r$ directly probes the corrugations of the contact discontinuity. Each spectrum corresponds to a SNR from \autoref{fig:SNR_timeline}, as indicated in the legend. \textbf{Inset:} $C(\ell)$ compensated by the $(a)$ spectrum, $C^{(a)}(\ell)$, to show which harmonics are growing with respect to the youngest SNRs in \autoref{fig:SNR_timeline}. The $C(\ell)$ admit to a power law $C(\ell) \propto \ell^{-8/3}$, which I show is consistent with 2D \citet{Kraichnan1967_two_dimensional_turbulence} turbulence, \autoref{eq:2D_turbulence}, that is self-generated on the thin shell, as discussed in \autoref{ssec:stuck_on_shell}. }
    \label{fig:spherical_harmonics}
\end{figure}

    \subsection{Measuring modes in the unstable thin shell}

    A detailed linear or nonlinear analysis of the shell, requiring spatio-temporal information about the layer, is beyond the scope of this Letter, where the focus is on linking turbulence to the $\baro$ generated in the layer and characterizing the layer itself. It is clear from the preceding analysis, and explicitly shown in the right sub-panels of \autoref{fig:vort_baro_slices}, that the layer is closely traced by intense $|\baro|$. Therefore, the pure $\baro$ spectrum, $\PB(k)$ (\autoref{eq:ps_b}), can be used to characterize the $k$-space structure of the layer. I plot $\PB(k)$ in the bottom panel of \autoref{fig:spectra}, ensemble averaged and normalized in the same way as the previous spectra. $\PB(k)\propto k^{3/2}$ across a broad range of $k$. In fluctuating dynamo theory, such a spectrum (the \citealt{Kazantsev1968} spectrum) is associated with an underlying folded geometry, where gradients pile up on small scales, producing folds in the layer down to the diffusion scale \citep{Schekochihin2004_dynamo,Kriel2022_kinematic_dynamo_scales,Kriel2025_SSD,Galishnikova2022_saturation_and_tearing,Kriel2025_nonlinear_dynamo}. For the \citet{Vishniac1983_overstability} overstability, driven by the mismatch in direction of the interior $\partial_r P$ on the shell and the exterior ram pressure across the shock, the fastest growing mode is of order the thin shell thickness, $k_{\rm max} \sim 1/\delta \xi_r$. Hence, with the spectrum peaking at high-$k$ this may indicate that $k_{\rm max}$ is close the grid-scale, which is expected without any explicit viscosity \citep{Badjin2021_SNR_instabilities}.

    In \autoref{fig:SNR_timeline} I plot the $\baro$ layer (bottom row) and $\baro$ sky (top row) for an observer at the center of the SNR, for four different SNRs with increasing $R_{\rm c}$ ($\sim$ age), illustrated in red in the bottom row (defined in next paragraph). All SNRs are taken from the same time realization and are expanding into a homogeneous background at the beginning of the global simulation (hence the $\delta \xi_r(\theta,\varphi)$ in the layer is not from interacting with an inhomogeneous background). High-$k_\perp$ modes and high-order harmonics grow as the SNRs age. By the time the SNRs expand to $\gtrsim 100\,\rm{pc}$ (far right; $R_{\rm c} \approx 50\,\rm{pc}$), the corrugations have become large cavities in the layer.

    To understand which modes are growing it is appropriate to calculate the spherical harmonics, $Y^{m}_{\ell}(\theta,\varphi)$, of $\delta \xi_r$ for the $\baro$ surface visualized in the bottom row of \autoref{fig:SNR_timeline}. I define $\xi_r(\theta,\varphi)$ as the radius of the $90^{\rm th}$ percentile isosurface of $\baro$ at each $(\theta,\varphi)$, which then allows me to construct $R_{\rm c} \equiv \xi_{r,\rm iso}$ by averaging over $\theta$ and $\varphi$ (shown in red), and finally $\delta\xi_r(\theta,\varphi)$ through \autoref{eq:displacement_vector}. I determine $a_{\ell m}$ by integrating $\xi_r = \sum_{\ell = 0}^{64}\sum_{m = -\ell}^\ell a_{\ell m} Y^{m}_{\ell}(\theta,\varphi)$, and then further sum over $m$ to probe the 1D $\ell$ harmonic spectrum, 
    \begin{align}\label{eq:spherical_ps}
        C(\ell) = \frac{1}{2}\sum_{m = -\ell}^{\ell} |a_{\ell m}|^2.         
    \end{align}

    In \autoref{fig:spherical_harmonics} I show the $C(\ell)$ uncompensated (main panel) and compensated by the $C(\ell)$ of SNRs $(a)$, $C^{(a)}(\ell)$ (inset panel). Firstly, all of the $C(\ell)$ admit to power laws, indicating that there is an entire spectrum of corrugations growing in approximately a self-similar manner such that $C(\ell)/C^{(a)}(\ell) \propto \ell^{\beta}$, $\beta > 0$. This is opposed to a single or small band of unstable modes, which would be a signature for an instability in the linear regime. The high-order harmonics (shown in the inset of \autoref{fig:spherical_harmonics}) have the largest $C(\ell)/C^{(a)}(\ell)$, but in late stages, SNRs $(d)$, the $\ell = 1$ and $\ell = 2$ dipole and quadrapole become $C(\ell)/C^{(a)}(\ell) \gg 1$, indicating that the SNRs have become deformed on large scales, consistent with radio observations of old SNRs \citep{Stafford2019_radio_obs_SNR}. There are no predictions from either the Vishniac, thermal, or Rayleigh Taylor instabilities that would provide a power law spectrum in harmonics. My interpretation is that this is a 2D turbulence cascade in the surface modes that form from the $\baro$, as discussed in \autoref{ssec:stuck_on_shell}. A corrugation interacting with the turbulence on the layer scales as $\delta \xi_r \sim u_t/k_\perp$, capturing the effect that larger $u_t$ will modify the corrugation more dramatically. For 2D \citet{Kraichnan1967_two_dimensional_turbulence} turbulence, $u_t^2 \propto k_\perp^{-2/3}$, hence,
    \begin{align}\label{eq:2D_turbulence}
        \delta \xi_r^2 \sim u_t^2/k_\perp^2 \sim k_{\perp}^{-8/3} \iff C(\ell) \propto \ell^{-8/3},
    \end{align}
    for $\ell \sim k_\perp R_{\rm c}$, which matches the data very well (shown with black dashed line in \autoref{fig:spherical_harmonics}). I interpret this as providing empirical evidence for the presence of 2D turbulence in unstable shell.
    
\section{Summary and conclusions}\label{sec:summary_and_conclusions}
    In this study I have constructed a unique dataset of three-dimensional SNRs ($n = 87$) extracted from multiphase gravitohydrodynamical simulations of SN-driven turbulence in a galactic disk cut-out \citep{Beattie2025_SLK41}. The simulations are parameterized (e.g., by mass, gravitational potential, cooling function, SN explosion rate) to match the large-scale ($L \gtrsim 10\,\rm{pc}$) properties of the Milky Way Galaxy. I present several key aspects of SN-driven turbulence that have not previously been revealed. All of my results emphasize the importance of the unstable contact discontinuity that develops once the SNR expands past its cooling radius, $R \approx 10-30\,\rm{pc}$ (with $t_{\rm age} \approx 10^4 - 10^5\,\rm{yr}$). I suggest that the unstable contact discontinuity is a key site for the generation of solenoidal turbulence in SN-driven systems, such as our Galaxy, and that the $k^{-3/2}$ slope observed in global SN-driven simulations originates at the SNR scale in the unstable layer.
    
    A summary of the key results are as follows:
    \begin{itemize}
        \item In \autoref{sec:generation_of_vorticity} I demonstrate that if baroclinicity, $\baro$, generates all of the incompressible turbulence at the SNR scale, then one can derive an analytical relation between the baroclinic–vorticity power cospectrum, $\PoB(k)$, which drives the turbulence, and the incompressible velocity-mode spectrum, $\Pus(k)$, \autoref{eq:baro_sol_relation}. This relation is strongly supported by my data (\autoref{fig:baro_enstrophy_spectra}), indicating the critical role of $\baro$ in sourcing the SN-driven turbulence. 
        \item In \autoref{app:baroclinic_source}, I show that during the onset of the SN-driven turbulence $\approx$ all of the $\baro$ is generated by the volume-poor, unstable contact discontinuity (\autoref{fig:baroclinc_source_1}), and in the steady state, $\gtrsim 70\%$  (\autoref{fig:baroclinc_source_2}). The classical phenomenology of shock-shock interactions, curved shocks, and corrugated shocks, is not supported empirically (\autoref{fig:baroclinc_source_3}). Hence, $\PoB(k)$, is dominated by the $\baro$ in the contact discontinuity both locally in the SNRs and globally in the disk cut-out simulations.
        \item Because the unstable layer drives a full spectrum of modes, SN-driven turbulence has no driving scale. Instead, it injects vorticity with $\PoB(k) \propto k^{3/4}$, peaking at high-$k$, for a $\Pus(k) \propto k^{-3/2}$ spectrum in incompressible velocity modes, \autoref{fig:baro_enstrophy_spectra}. Both spectra lack a first principles derivation, but it is shown clearly in the data, which I defer for future work utilizing local simulations of the layer.
        \item Based upon the geometric nature of $\baro$, the modes that are generated are surface modes that span the tangent plane of the unstable contact discontinuity, which I show in \autoref{ssec:stuck_on_shell}. However, the radial velocities from the SNRs couple to the surfaces modes via vortex stretching, and can efficiently eject or shed them into the surrounding medium. From the vortex stretching equation, \autoref{eq:vortex_stretching}, I show that the criterion is $t_{\rm nl}/t_{\rm 3D} > 1$, where $_{\rm nl}$ is the nonlinear timescale of the modes on the surface and $t_{\rm 3D}$ is the timescale of the vortex stretching into the third dimension, \autoref{eq:confinment_times}. I show that young SNRs efficiently shed their surface modes, and old SNRs do not, making SNRs with $t_{\rm age} \sim t_{\rm cool}$ the most efficient generators of turbulence.
        \item The unstable layer exhibits a $k^{3/2}$ power spectrum in $\baro$ (bottom panel, \autoref{fig:spectra}) and a $C(\ell) \propto \ell^{-8/3}$ spectrum in spherical harmonics of the corrugations (\autoref{eq:spherical_ps}). This a signature of a highly folded geometry (well known from small-scale dynamo theory). I conjecture that the unstable modes are seeded by an interface instability \citep[e.g.,][]{Vishniac1983_overstability,MacLow1993_overstability,Badjin2021_SNR_instabilities}, rather than from interactions with an inhomogeneous medium. I show the $C(\ell) \propto \ell^{-8/3}$ spectrum can be predicted by the 2D \citet{Kraichnan1967_two_dimensional_turbulence} turbulence model, providing evidence that the power law harmonic spectrum is generated from surface turbulence on the contact discontinuity.
        \item I show the incompressible turbulence spectrum generated by $\baro$ is $\Pus(k) \propto k^{-3/2}$ (top panel, \autoref{fig:spectra}), which develops on the local SNR scale. Simulations of SN-driven turbulence in galactic disks also show a $\Pus(k) \propto k^{-3/2}$ spectrum \citep{Padoan2016_supernova_driving,Beattie2025_SLK41}, extending from the disk into the galactic winds \citep{Connor2025_cascading_from_the_winds}. Therefore, I suggest that the $\Pus(k) \propto k^{-3/2}$ spectrum generated in the unstable layer couples to compressible modes generated by the SNRs, imprinting itself on the largest scales of the galactic turbulence cascade. This provides a striking example of how small-scale instabilities may be able to shape the large-scale properties of galaxies and defines a new phenomenology for SN-driven turbulence.
    \end{itemize}
    
    A number of outstanding problems remain. (1) Based on \citet{Beattie2025_SLK41} and \citet{Connor2025_cascading_from_the_winds}, the $\Pus(k) \propto k^{-3/2}$ spectrum generated by the unstable contact discontinuity appears to survive transport from the galactic disk into the winds. However, galactic shear could deform the growing modes, altering the spectrum and preventing it from being perfectly imprinted on the largest scales. The survival of the spectrum under shear must be tested in future simulations. (2) An inhomogeneous or turbulent upstream medium inevitably perturbs the shock and introduces additional structure into the thin radiative cooling shell. Consequently, the driving spectrum $\PoB(k)$ may vary between individual SNRs. However, this may not erase the signature of the unstable layer. Quantities that ensemble average over SNRs, like the global velocity power spectrum, may retain a systematic imprint of the unstable layer because $\Pus(k)$ is the sum over the spatial variations, allowing the spectrum to retain the systematic contribution from the unstable layer at the level of the velocity dispersion. (3) The role of magnetic fields will be crucial in further developing this new SN-driven turbulence phenomenology, since they are generated proportionally with $\baro$ through the Biermann battery, when the electron pressure is not spatially uniform \citep{Biermann1950_battery,Beattie2025_SLK41}. A detailed study of the Biermann effect, $\partial_t \bm{B} \propto \baro$, and the inevitable dynamo is deferred to future work, when this problem setup is ported to the GPU-enabled \textsc{AthenaK} code \citep{Stone2024_athenaK}. This will enable substantially higher-resolution calculations and a more direct characterization of the two-dimensional surface modes generated by $\baro$.
        
\section*{\textbf{Acknowledgments}}
    I thank the reviewer, Mordecai-Mark Mac Low, for suggesting additional calculations to strengthen the arguments presented in this work, as well as both their detailed discussion about the overstability versus the nonlinear thin shell instability, and the detailed linguistic comments that helped make the manuscript clearer and more precise. I thank Isabelle Connor, Enrico Ramirez-Ruiz, Ish Kaul, Chris Thompson, Michael Grehan, Chang-Goo Kim, Chris Bambic, Norm Murray, Ralf Klessen, Amitava Bhattacharjee, Bart Ripperda and the CITA plasma-astro group for many enlightening discussions. I acknowledge compute allocations rrg-ripperda and rrg-essick from the Digital Research Alliance of Canada, which supported parts of this project; funding from the Natural Sciences and Engineering Research Council of Canada (NSERC, funding reference number 568580); support from NSF Award 2206756; and high-performance computing resources provided by the Leibniz Rechenzentrum and the Gauss Center for Supercomputing (grants pn76gi, pr73fi, and pn76ga).

    \software{I use \textsc{ramses} \citep{Teyssier2002_ramses} for all of the simulations. Data analysis and visualization software used in this study: \textsc{C++} \citep{Stroustrup2013}, \textsc{numpy} \citep{Oliphant2006,numpy2020}, \textsc{numba}, \citep{numba:2015}, \textsc{matplotlib} \citep{Hunter2007}, \textsc{cython} \citep{Behnel2011}, \textsc{visit} \citep{Childs2012}, \textsc{scipy} \citep{Virtanen2020}, \textsc{scikit-image} \citep{vanderWalts2014}, \textsc{cmasher} \citep{Velden2020_cmasher}, \textsc{joblib}\citep{joblib}, \textsc{pyfftw}\citep{2021ascl.soft09009G}, \textsc{healpy}\citep{Healpy2019}, \textsc{healpix}\citep{HealPix2005}}

\appendix

\section{Examples of identified SNR shown in the global frame}\label{app:clusters}
\begin{figure*}
        \centering
        \includegraphics[width=\linewidth]{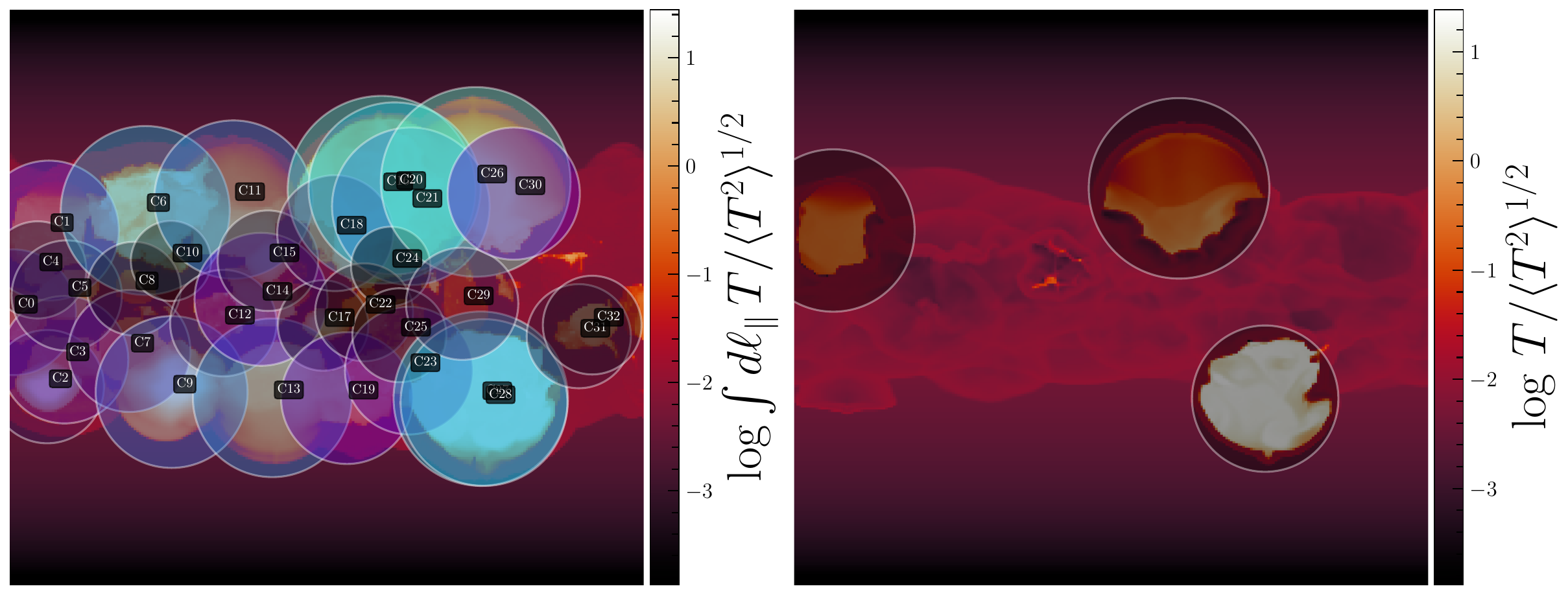}
        \caption{\textbf{Identified SNRs in disk cut-out simulations.} \textbf{Left:} An example showing 33 supernova remnants identified at early times in the simulations from \citet{Beattie2025_SLK41}. The background field is the logarithmic line-of-sight temperature, normalized by the root-mean-squared temperature of the full domain. Each label marks the projected geometric center of an SNR, which I use to extract a surrounding $128^3$-cell region for the local analysis presented in the main text. \textbf{Right:} the same as the left plot but for the a slice through the middle of the out-of-page axis.}
        \label{fig:clusters}
    \end{figure*}

    Using the method described in \autoref{sec:sims}, I extract nearly 100 SNRs at early times from the simulations in \citet{Beattie2025_SLK41}. Here I provide an example illustrating how these SNRs appear when grouped together in the global simulation. In the left panel of  \autoref{fig:clusters}, I show 33 SNRs extracted from a single time realization. The background field displays the logarithmic line-of-sight (LOS) integrated temperature, normalized by its root-mean-squared value. Each label identifies an individual SNR and marks its projected geometric center. I use the geometric center to define a surrounding $128^3$-cell patch for the local analysis presented in this study. In the right panel I show the same, but sliced through the middle of the orthogonal coordinate to the disk, instead of projecting.

    \begin{figure*}
        \centering
        \includegraphics[width=\linewidth]{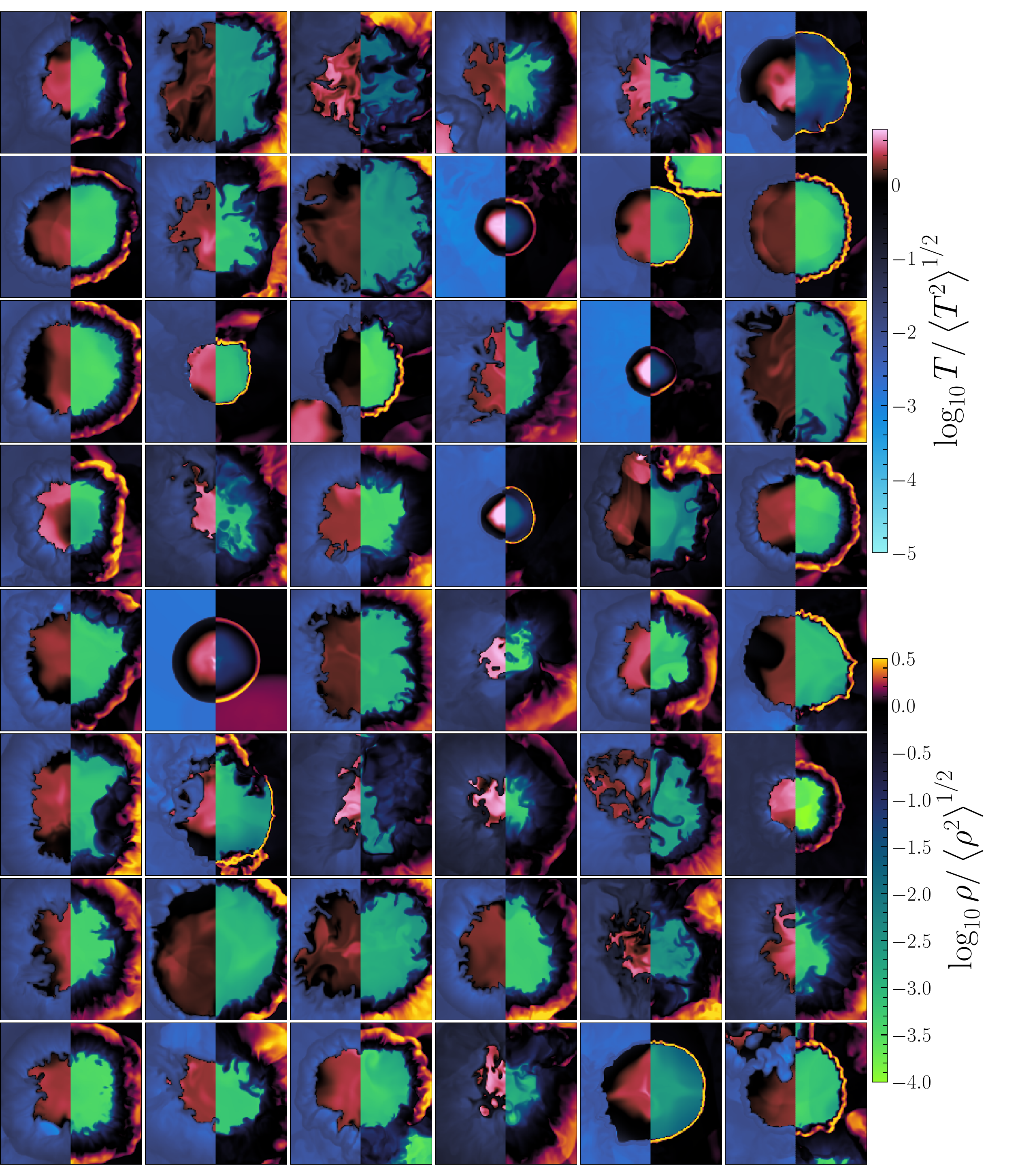 }
        \caption{The same as \autoref{fig:vort_baro_slices} but for the logarithmic, root-mean-square normalized temperature, $T/\langle T^2 \rangle^{1/2}$ (left in each panel), and mass density, $\rho /\langle \rho^2 \rangle^{1/2}$ (right in each panel). The SNRs exhibit diverse morphologies depending on their age (radii). Young SNRs are nearly spherical, with only high-$k$ modes present in the thin density interface, the contact discontinuity, between the hot (red) and warm (blue) plasma. Older SNRs develop highly fractal interfaces, where denser, cooler plasma penetrates deep into the remnant.}
        \label{fig:temperature_slices}
    \end{figure*}

    I show a random sample of 48 SNRs (the same as in \autoref{fig:vort_baro_slices}) in \autoref{fig:temperature_slices}, with two-dimensional slices of temperature $\log T / \left\langle T^2 \right\rangle^{1/2}$ (left) and mass density $\log \rho / \left\langle \rho^2 \right\rangle^{1/2}$ (right). The SNRs exhibit diverse morphologies in both $T$ and $\rho$. Young SNRs are nearly spherical, with a thin (yellow) $\rho$ shell at the cooling radius that forms at the boundary between the hot (red) and warm (blue) plasma, i.e., the contact discontinuity. Older SNRs develop fractal interfacial surfaces in $\rho$ and $T$, along with more deformed overall contact discontinuity cross-sections. The fractal $T$ interface is reminiscent of the fingers produced by the nonlinear Rayleigh–Taylor instability, whereas the unstable $\rho$ layer more closely resembles a thin-shell instability \citep{Vishniac1983_overstability,MacLow1993_overstability,McLeod2013_simulations_of_the_nonlinearVish}. Previous studies have found a qualitatively similar thin shell instabilities in expanding SNR in homogeneous media, as long as they have explicit cooling and the expansion is not purely adiabatic in (Figure~3 in \citealt{Kim2015_SNR_expansion}; Figure~2, HC simulation in \citealt{Walch2015_SNR}; and Figure~14 in \citealt{Guo2024_high_res_SN}).

\section{The source of incompressible turbulence: curved shocks, shock interactions, and the unstable contact discontinuity in SNR}\label{app:baroclinic_source}
    \begin{figure*}
        \centering
        \includegraphics[width=\linewidth]{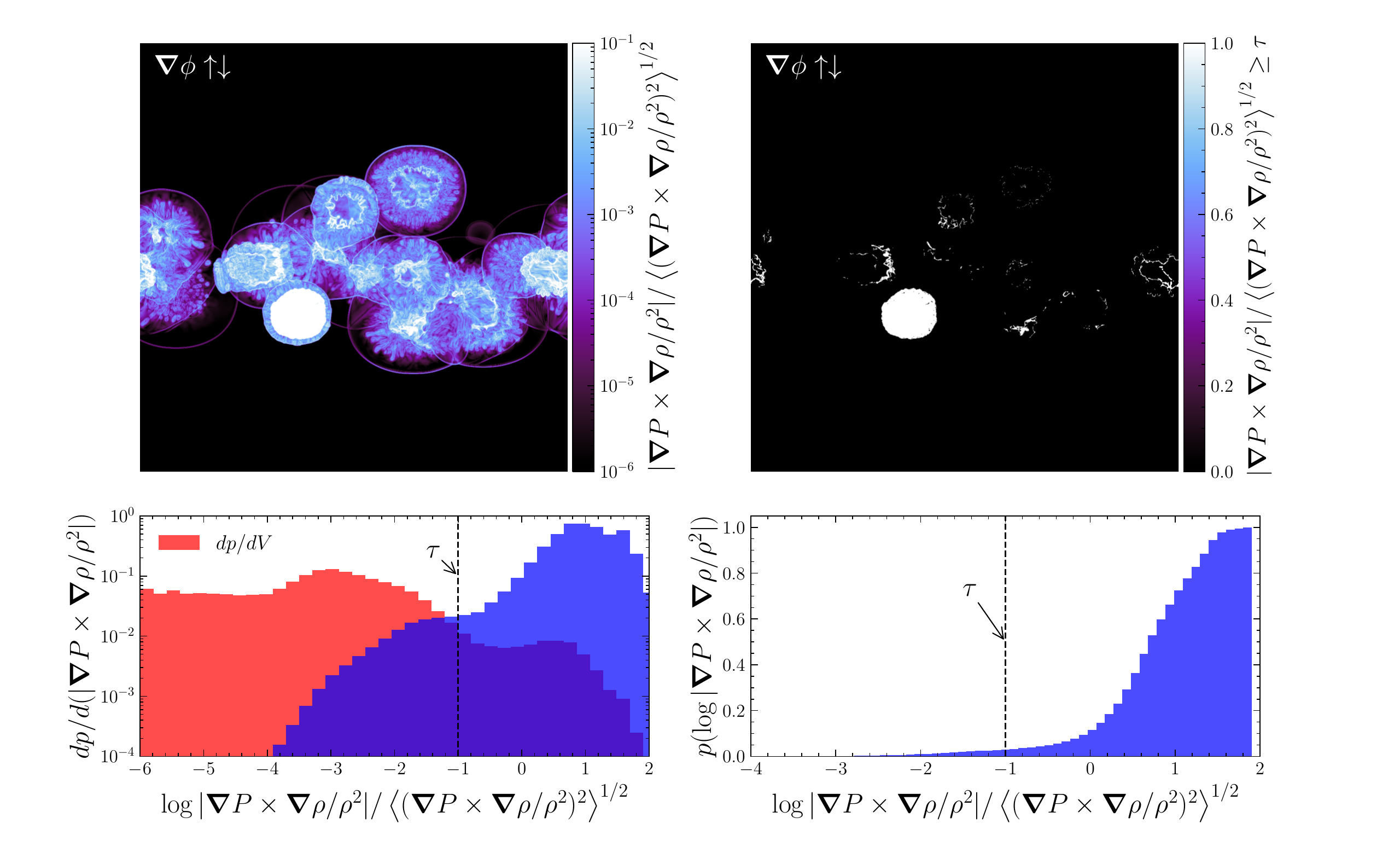}
        \caption{\textbf{Global baroclinicity statistics at the onset of SN-driven turbulence.}
        \textbf{Top-left:} Slice of $|\baro|$, normalized by the volume-integrated root-mean-squared value, shown in a plane parallel to the gravitational potential gradient, $\bnab\phi$. \textbf{Bottom-left:} The $|\baro|$-weighted (blue) and volume-weighted (red) PDFs of $\log|\baro|$. These demonstrate that baroclinic generation is concentrated in the high-$|\baro|$, low-volume tail of the volume-weighted distribution. A threshold $\tau = 0.1\,\langle (\baro)^2 \rangle^{1/2}$ is indicated as the approximate point where the $|\baro|$-weighted PDF begins to rise.
        \textbf{Top-right:} Same slice as in the top-left panel, but retaining only regions with $|\baro|>\tau$, highlighting the structures responsible for the majority of baroclinic production. As discussed here and in \citet{Beattie2025_SLK41}, these correspond to the unstable thin shells that form near the cooling radius of SNRs (the unstable contact discontinuity).
        \textbf{Bottom-right:} Cumulative distribution function of $|\baro|$, showing that $\partial_t|\bom| \propto |\baro|$ is overwhelmingly dominated by regions with $|\baro|/\langle (\baro)^2 \rangle^{1/2} > \tau$, i.e. the high-baroclinicity structures isolated in the top-right panel.}
        \label{fig:baroclinc_source_1}
    \end{figure*}

    In this section, I demonstrate that the unstable contact discontinuity is the leading-order source of baroclinicity in the simulations from \citet{Beattie2025_SLK41}. The approach is intentionally simple. I first compute $\baro$ from the full simulation domain, and then evaluate both the volume-weighted probability density function (PDF), $dp/dV$, and the $|\baro|$-weighted PDF, $dp/d(|\baro|)$, along with the corresponding cumulative distribution function (CDF) $p(X' \leq X) = \int^X_{-\infty} dX' dp(X')/dX'$. These diagnostics are computed for two time realizations: (1) an early epoch, before the turbulence reaches statistical steady state and when many SNRs still appear pristine (this corresponds to the period in which I extract SNRs for isolated analysis); and (2) the later, steady-state turbulent phase. From \citet{Beattie2025_SLK41}, these times correspond to $t = 4t_0$ and $t = 30t_0$, respectively, matching the first two columns of Figure~4 and Figure~11 in that study. Using the PDFs, CDFs, and visualizations of slices of $|\baro|$, I identify which physical structures dominate $|\baro|$ and therefore dominate $\partial_t\bom \propto k \partial_t \mathbf{u}_s$ in the SN-driven turbulent state.
    
    In \autoref{fig:baroclinc_source_1}, I present the results for the early-time realization in which SNRs expand into an effectively homogeneous medium, prior to the development of steady-state turbulence. This time frame closely resembles the conditions under which I extract the SNRs for isolating internal instabilities from background inhomogeneities and large-scale shock interactions. The top-left panel shows a slice of $|\baro|$, normalized by its root-mean-squared (rms) value. The bottom-left panel shows the volume-weighted (red) and $|\baro|$-weighted (blue) PDFs. Their comparison immediately reveals that most of the baroclinicity (and hence most of $\partial_t |\bom|$) is generated by extremely low-volume-filling structures residing in the high-$|\baro|$ tail—well above the rms value—with the onset occurring at roughly 1\% of the rms. To quantify this, I compute the CDF and mark a threshold $\tau = 0.1\langle (\baro)^2\rangle^{1/2}$ on both the PDF and CDF panels. From the CDF, it is evident that almost all of the $|\baro|$ originates from regions where $|\baro|/\langle (\baro)^2 \rangle^{1/2} \ge \tau$.
    
    Applying this threshold to the spatial data—masking out all regions with $|\baro|/\langle (\baro)^2 \rangle^{1/2} < \tau$ yields the top-right panel of \autoref{fig:baroclinc_source_1}. This visualization makes clear that essentially all of the baroclinicity is generated directly within the unstable layers identified in \citet{Beattie2025_SLK41}. Thus, at early times, the low-volume-filling unstable shells at the cooling radius of each SNR are the principal source of baroclinicity and therefore the dominant engine of the turbulence in the simulations. These unstable layers drive an entire spectrum of incompressible modes that follow the relationship derived in \autoref{eq:baro_sol_relation}.

    \begin{figure*}
        \centering
        \includegraphics[width=\linewidth]{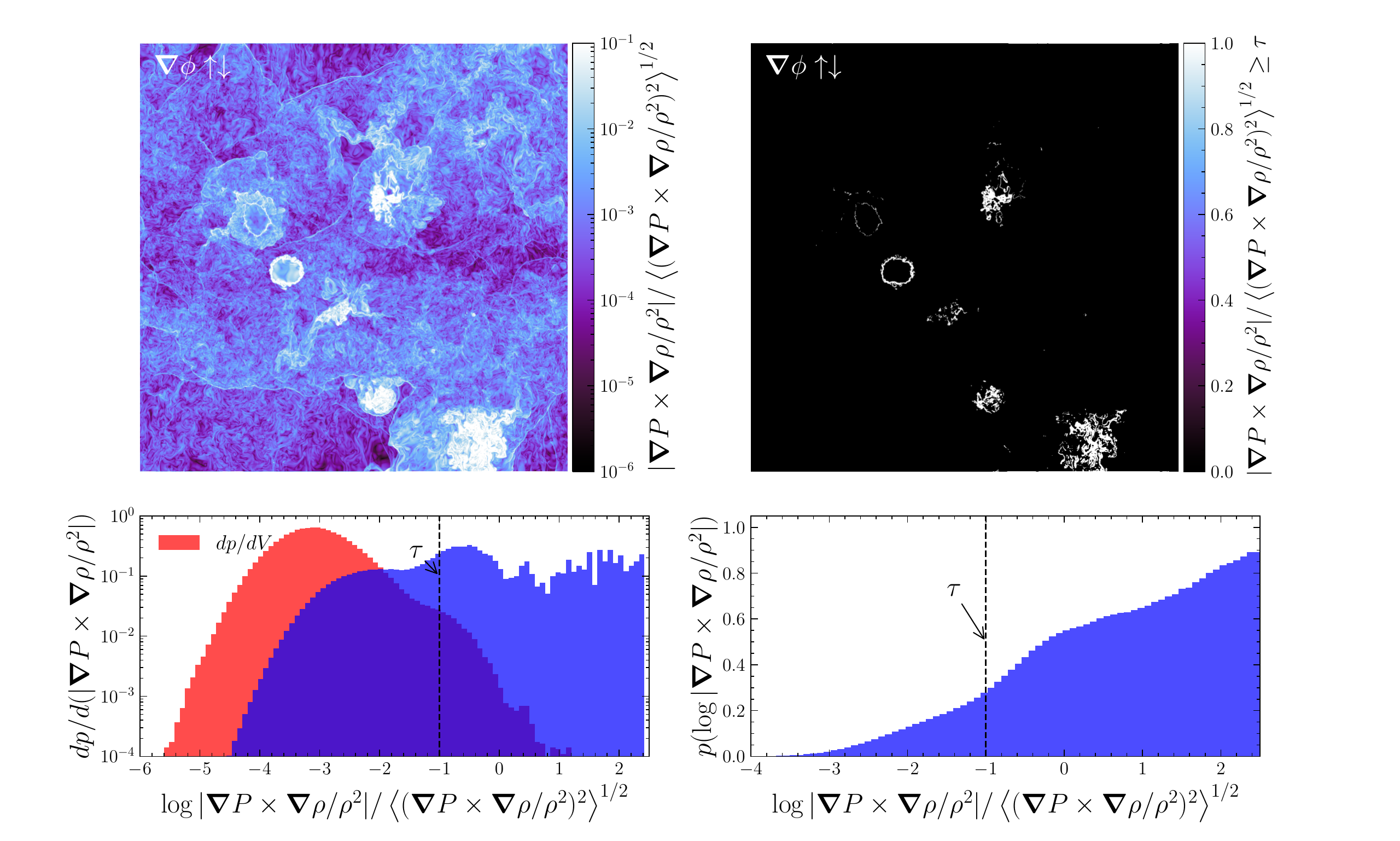}
        \caption{\textbf{Global baroclinicity statistics in the steady-state SN-driven turbulence.}
        The same diagnostics as in \autoref{fig:baroclinc_source_1}, but now evaluated in the statistically steady state of SN-driven turbulence. The results demonstrate that the same mechanism responsible for initially driving the turbulence continues to dominate its maintenance: the low-volume, unstable layer between the hot and warm plasma in each SNR produces the vast majority of the baroclinicity (and hence incompressible turbulence). Regions with $|\baro|/\langle(\baro)^2\rangle^{1/2} >\tau$ contain $\approx$ 70\% of the total baroclinic amplitude, confirming that the unstable layers remain, to leading order, the primary engines of the incompressible turbulence even in steady state.}
        \label{fig:baroclinc_source_2}
    \end{figure*}

    \begin{figure*}
        \centering
        \includegraphics[width=0.7\linewidth]{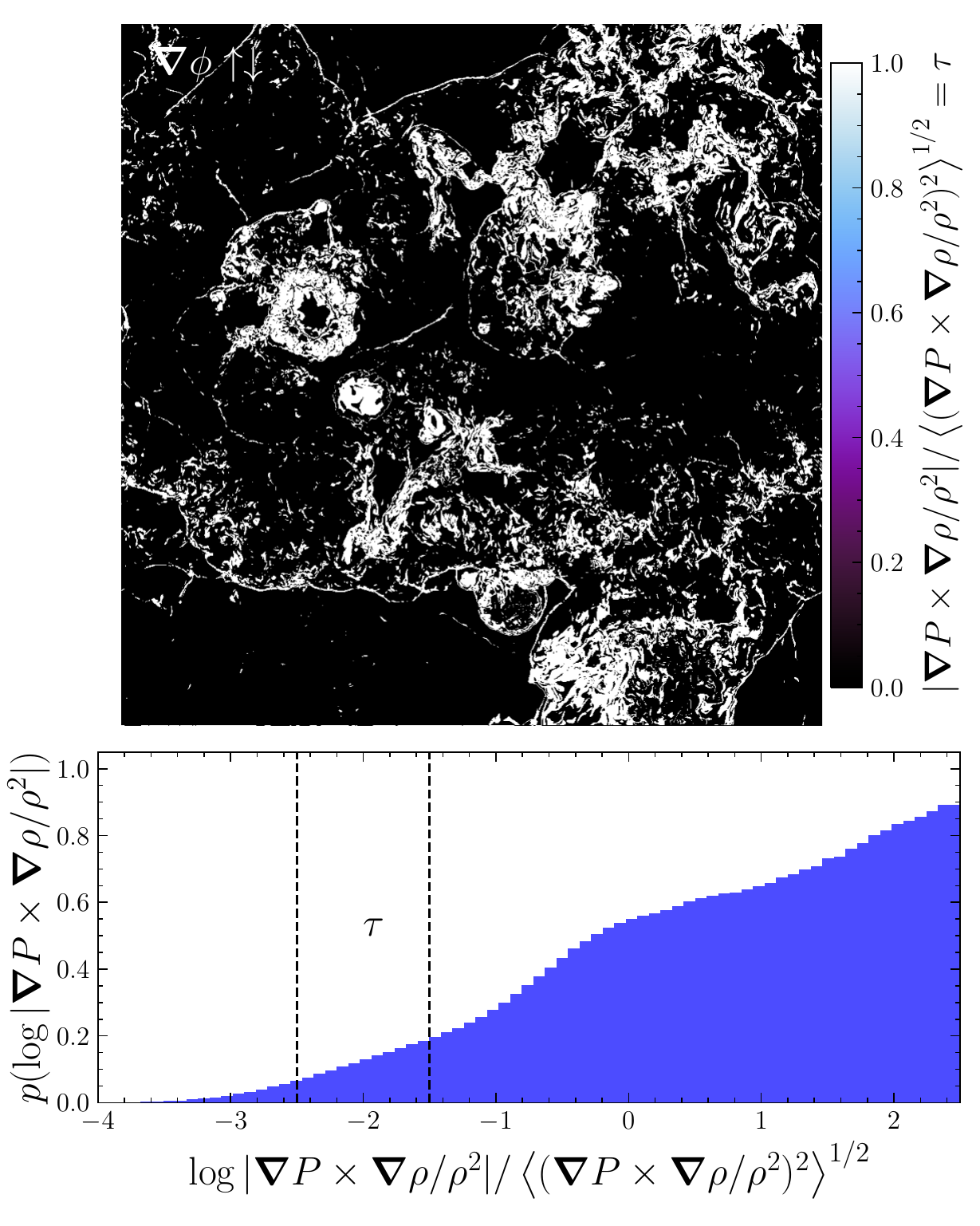}
        \caption{\textbf{Baroclinicity from shock interactions and curved shocks in steady-state SN-driven turbulence.}
        Same diagnostics as in the right panels of \autoref{fig:baroclinc_source_1}, but using a threshold $\tau$ chosen specifically to isolate curved shocks and shock–shock interactions. With this selection, these features contribute only $\approx $10\% of the total $|\baro|$, and only at relatively low amplitudes (corresponding to slower vorticity generation rates, $\baro \sim 1/t_0^2$). Thus, baroclinic production in the steady state is strongly dominated by the unstable contact discontinuities highlighted in \autoref{fig:baroclinc_source_1} and \autoref{fig:baroclinc_source_2}.}
        \label{fig:baroclinc_source_3}
    \end{figure*}

    I now repeat the analysis for the statistically steady state of SN-driven turbulence. \autoref{fig:baroclinc_source_2} shows the same diagnostics as \autoref{fig:baroclinc_source_1}, but for a time realization within the steady state. From the top-left panel, the medium is clearly highly inhomogeneous: structures fill the disk and extend into the galactic wind. Despite this complexity, the PDFs, CDFs, and the slice masked by $|\baro|/\langle (\baro)^2 \rangle^{1/2} \ge \tau$ demonstrate that the basic result is unchanged. The baroclinicity, and therefore the incompressible turbulent driving, continues to arise overwhelmingly from extremely low-volume-filling structures. When these structures are isolated using the same threshold $\tau$, they again correspond to the contact discontinuities, now contributing $\gtrsim$ 70\% of the total $|\baro|$ at this $\tau$. This is entirely consistent with the phenomenology proposed in \citet{Beattie2025_SLK41} and in this study.

    To complete this appendix section, I vary $\tau$ to identify the regions in the CDF associated with shock–shock interactions and curved and corrugated shocks. These features have been proposed as primary contributors to baroclinicity in previous theoretical work \citep{Elmegreen2004,Balsara2004_SNe_driven_turb,Kapyla2018_vorticity_helicity_ISM}. I find that thresholds in the range $-2.5 \le \log\left(\tau/\langle (\baro)^2 \rangle^{1/2}\right) \le -1.5$ isolate numerous shell-like structures on both small and large scales, as shown in \autoref{fig:baroclinc_source_3}. I therefore identify these as the baroclinic signatures of shocks. For these values of $\tau$—indicated by vertical lines in the CDF—the shock–shock interactions and curved shocks contribute only $\approx$ 10\% of the total $|\baro|$. Thus, to leading order (i.e., for all $|\baro|/\langle (\baro)^2 \rangle^{1/2} \ge 0.1\langle (\baro)^2 \rangle^{1/2}$, which accounts for over 70\% of the total baroclinicity; see \autoref{fig:baroclinc_source_2}), it is the unstable contact discontinuities—not shock interactions or curved shocks—that sustain the incompressible turbulence.
    
    This explains why idealized turbulence-box simulations, even when driven purely solenoidally, fail to produce as much energy in the solenoidal velocity component as SN-driven turbulence \citep{Federrath2010_solendoidal_versus_compressive,Padoan2016_supernova_driving,Beattie2025_SLK41}: they lack the inherently and strongly baroclinic thin shells that naturally arise from the geometry of expanding, cooling SNRs. These are the engines of SN-driven turbulence.
    
\section{Additional spectra}
\subsection{Incompressible velocity mode and vorticity spectrum correspondence}\label{app:spectrum_vort}
    In \autoref{eq:baro_sol_relation} I derived a relation between the vorticity–baroclinic interaction spectrum, $\PoB(k)$, and the incompressible mode spectrum, $\Pus(k)$. This relied on the fact that the vorticity spectrum, $\Po(k)$, can be written simply as $\Po(k) \sim k^2 \Pus(k)$. \citet{Connor2025_cascading_from_the_winds}, in their Appendix, showed that this relation also holds for the compressible mode $\u_c$, where $|\bnab \times \u_c|=0$, with the counterpart relation $\mathcal{P}_{\u_c}(k) \sim k^2 \mathcal{P}_{\bnab \cdot \u}(k)$, exactly as one would hope and expect. For completeness, I confirm the incompressible–vorticity spectrum relation here. In the left panel of \autoref{fig:vrt_spectrum} I show $\Pus(k)$ and $\Po(k)/k^2$ (with the same normalizations as \autoref{fig:spectra}), illustrating good agreement between the two power spectra, $\Pus(k) \sim k^2\Po(k)$, across all long wavelengths ($k\ell_0/2\pi \lesssim 10$). These are the modes that are not influenced by numerical viscosity in these simulations. Since $\Pus \propto k^{-3/2}$ (as shown in \autoref{fig:spectra}), the vorticity follows $\Po(k) \propto k^{1/2}$.

\subsection{Baroclinic - vorticity cospectrum (the turbulence driving spectrum)}
\begin{figure}
    \centering
    \includegraphics[width=\linewidth]{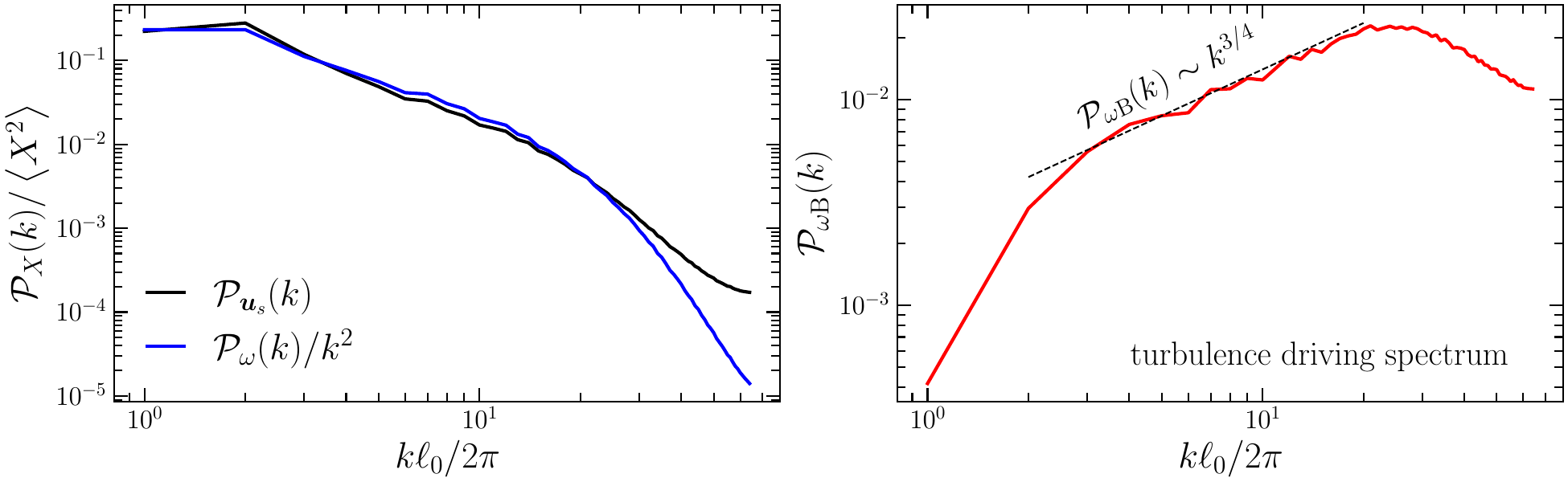}
    \caption{\textbf{Left:} The same as \autoref{fig:spectra}, but for the incompressible-velocity mode spectrum, $\Pus(k)$ (black), and the vorticity spectrum, $\Po(k)$ (blue). $\Po(k)$ is compensated by $k^2$ to test the canonical prediction $\Pus(k) \sim k^2\Po(k)$. The spectra are averaged over all SNRs. For modes with wavelengths larger than the diffusion-dominated scales ($k\ell_0/2\pi \lesssim 10$), the two spectra correspond almost perfectly, as expected, implying $\Po(k) \propto k^{1/2}$. \textbf{Right:} the baroclinicity and vorticity cospectrum \autoref{eq:ps_ob}, $\PoB(k)$, i.e., the spectrum that probes the flux interaction between $\baro$ and $\bom$. The $\PoB(k) \propto k^{3/4}$ relation, showing a perfect match to the spectrum, comes from my prediction for a $\Pus(k) \propto k^{-3/2}$ spectrum that is sourced completely from $\bom\cdot(\baro)$, \autoref{eq:baro_sol_relation}.}
    \label{fig:vrt_spectrum}
\end{figure}

    From \autoref{eq:baro_sol_relation} I predict that there is a scaling between the baroclinic - vorticity cospectrum, $\PoB(k)$, \autoref{eq:ps_ob}, and the incompressible mode spectrum, $\Pus(k)$, \autoref{eq:ps_sol}, assuming that $\Pus(k)$ is completely sourced by the flux from $\baro$ into $\bom$. For a $\Pus(k) \propto k^{-3/2}$ spectrum, as I find in this paper, \autoref{fig:spectra}, and in previous SN-driven turbulence papers, probing larger scales \citep{Padoan2016_supernova_driving,Beattie2025_SLK41,Connor2025_cascading_from_the_winds}, using \autoref{eq:baro_sol_relation} I derive that $\PoB(k) \propto k^{3/4}$. In the left panel of \autoref{fig:vrt_spectrum} I show that $\PoB(k) \propto k^{3/4}$ matches the cospectrum perfectly. This means the predicted driving spectrum for SN-driven turbulence peaks at high-$k$, and the energy in the long wavelengths falls off with a power-law, and then even steeper at the longest wavelengths in the\ domain (note that $\ell_0 = 125\,\rm{pc}$). Hence the importance of the inverse cascade, as outlined by \citet{Beattie2025_SLK41}, allowing for the turbulence, produced on small-scales, to reach beyond the gaseous-scale height of the disk, as shown in \citet{Connor2025_cascading_from_the_winds}.

\section{Timescales for surface modes versus 3D vortex shedding}\label{app:confinement_time}
\begin{figure}
    \centering
    \includegraphics[width=0.6\linewidth]{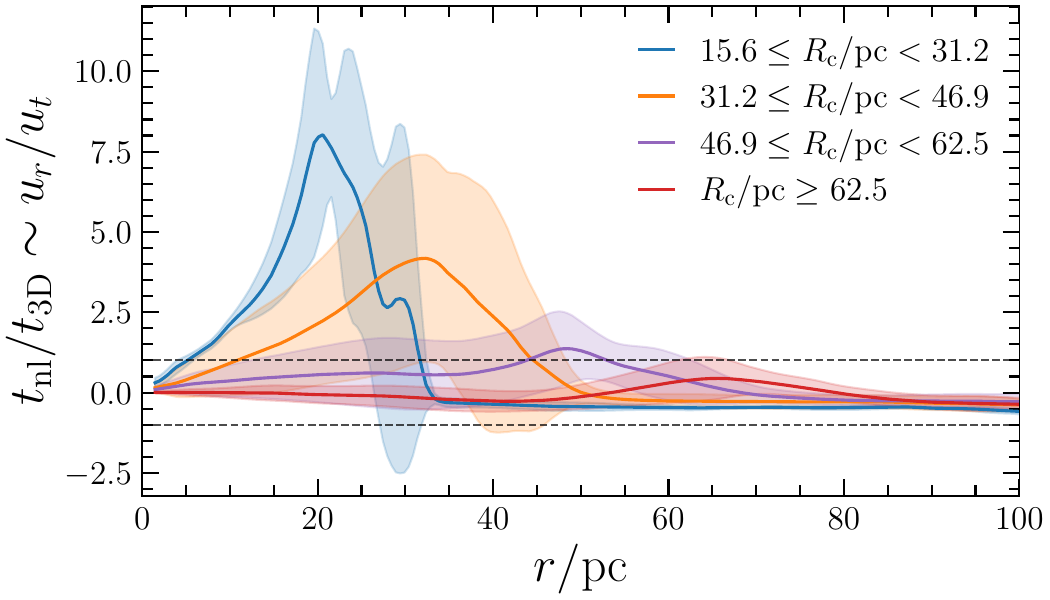}
    \caption{The ratio between the 2D nonlinear timescale, $t_{\rm nl} = 1/(k_\perp u_t)$, and the 3D vortex stretching timescale, $t_{\rm 3D} \sim 1/(k_{\perp}u_r)$, where $k_\perp$ are modes on the surface of the contact discontinuity, $u_t$ is the turbulent component of the velocity in the transverse plane to the surface, and $u_r$ is the radial velocity (a compressible mode), penetrating through the unstable layer (discussed in \autoref{ssec:stuck_on_shell}). Each line represents a different population of SNR, binned by the radius of the contact discontinuity, $R_c$ ($\sim$ age). The $\pm1$ is indicated with the black, dashed lines. For $t_{\rm nl} /t_{\rm 3D} > 1$, the modes generated on the surface shed into the surrounding medium through vortex stretching. For $t_{\rm nl} /t_{\rm 3D} < 1$, the modes are confined to interact (and potentially cascade) on the surface. Old SNR may become confined due to the decaying $u_r$, whereas young SNR can efficiently radiate the turbulence that develops at the contact discontinuity.}
    \label{fig:confinement_time}
\end{figure}

    I derive in \autoref{ssec:stuck_on_shell} the criterion for surfaces modes on the contact discontinuity to be stretched into the third dimension through the $u_r$ velocities coming from the SN explosion. Directly from the vortex stretching equation, the criterion for small corrugations, $\delta\xi_r \ll R_c k_\perp$, is $t_{\rm nl} / t_{\rm 3D} > 1$, where $t_{\rm nl} \sim 1/(k_\perp u_t)$ is the nonlinear timescale of the surface modes and $t_{\rm 3D} \sim 1/(k_\perp u_r)$ is the timescale for stretching the modes and generating $\omega_r$. In \autoref{fig:confinement_time} I directly measure $t_{\rm nl} / t_{\rm 3D}$ for the population of SNR I analyze in this study. 
    
    First I take $\Exp{u_r}{\theta,\varphi}$, $\Exp{u_t^2}{\theta,\varphi}^{1/2}$ and $\rho$ profiles for every SNR, which I use to construct $t_{\rm nl} / t_{\rm 3D}$. I use the root-mean-squared $u_t$ because it is for the nonlinear timescale. By inspecting $\rho$, I find the peak in $t_{\rm nl} / t_{\rm 3D} \sim u_r / u_t$ occurs at the contact discontinuity, so one can think of the peak of $t_{\rm nl} / t_{\rm 3D}$ and $R_{\rm c}$ interchangeable. To see if there is a difference between young and old SNR, I bin the population by $R_{\rm c}$ into the four groups I have annotated in the legend. The key result is that young SNR in my population are in the $t_{\rm nl} / t_{\rm 3D} > 1$ regime (I annotate $\pm1$ with black, dashed lines), allowing them to shed the surface turbulence. On the other hand, old SNR are in the $t_{\rm nl} / t_{\rm 3D} < 1$ regime, where 2D modes interact on the surface on shorter timescales than they are shed. 

\bibliography{Mar24,james_bib}

@ARTICLE{Teyssier2002_ramses,
       author = {{Teyssier}, R.},
        title = "{Cosmological hydrodynamics with adaptive mesh refinement. A new high resolution code called RAMSES}",
      journal = {\aap},
         year = 2002,
        month = apr,
       volume = {385},
        pages = {337-364},
          doi = {10.1051/0004-6361:20011817},
archivePrefix = {arXiv},
       eprint = {astro-ph/0111367},
 primaryClass = {astro-ph},
       adsurl = {https://ui.adsabs.harvard.edu/abs/2002A&A...385..337T}
}

@software{joblib,
title = {Joblib: running Python functions as pipeline jobs},
author = {{Joblib Development Team}},
year = {2020},
url = {https://joblib.readthedocs.io/},
}

@ARTICLE{Martizzi2016,
       author = {{Martizzi}, Davide and {Fielding}, Drummond and {Faucher-Gigu{\`e}re}, Claude-Andr{\'e} and {Quataert}, Eliot},
        title = "{Supernova feedback in a local vertically stratified medium: interstellar turbulence and galactic winds}",
      journal = {The Monthly Notices of The Royal Astronomical Society},
         year = 2016,
        month = jul,
       volume = {459},
       number = {3},
        pages = {2311-2326},
          doi = {10.1093/mnras/stw745},
archivePrefix = {arXiv},
       eprint = {1601.03399},
 primaryClass = {astro-ph.GA},
       adsurl = {https://ui.adsabs.harvard.edu/abs/2016MNRAS.459.2311M}
}

@ARTICLE{Beattie2025_SLK41,
       author = {{Beattie}, James R. and {Kolborg}, Anne Noer and {Ramirez-Ruiz}, Enrico and {Federrath}, Christoph},
        title = "{So Long Kolmogorov: The Forward and Backward Turbulence Cascades in a Supernovae-driven, Multiphase Interstellar Medium}",
      journal = {\apj},
         year = 2025,
        month = dec,
       volume = {994},
       number = {2},
          eid = {193},
        pages = {193},
          doi = {10.3847/1538-4357/ae07cd},
archivePrefix = {arXiv},
       eprint = {2501.09855},
 primaryClass = {astro-ph.GA},
       adsurl = {https://ui.adsabs.harvard.edu/abs/2025ApJ...994..193B}
}

@ARTICLE{MacLow1993_overstability,
       author = {{Mac Low}, Mordecai-Mark and {Norman}, Michael L.},
        title = "{Nonlinear Growth of Dynamical Overstabilities in Blast Waves}",
      journal = {\apj},
         year = 1993,
        month = apr,
       volume = {407},
        pages = {207},
          doi = {10.1086/172506},
       adsurl = {https://ui.adsabs.harvard.edu/abs/1993ApJ...407..207M}
}

@ARTICLE{Grehan2025_num_diffusion,
       author = {{Grehan}, Michael P. and {Ghosal}, Tanisha and {Beattie}, James R. and {Ripperda}, Bart and {Porth}, Oliver and {Bacchini}, Fabio},
        title = "{Comparison of magnetic diffusion and reconnection in ideal and resistive relativistic magnetohydrodynamics, ideal magnetodynamics, and resistive force-free electrodynamics}",
      journal = {\prd},
         year = 2025,
        month = sep,
       volume = {112},
       number = {6},
          eid = {063046},
        pages = {063046},
          doi = {10.1103/8xf2-x2nq},
archivePrefix = {arXiv},
       eprint = {2503.20013},
 primaryClass = {astro-ph.HE},
       adsurl = {https://ui.adsabs.harvard.edu/abs/2025PhRvD.112f3046G}
}

@ARTICLE{Badjin2021_SNR_instabilities,
       author = {{Badjin}, Dmitry A. and {Glazyrin}, Semyon I.},
        title = "{Physical and numerical instabilities of radiatively cooling shocks in turbulent magnetized media}",
      journal = {\mnras},
         year = 2021,
        month = oct,
       volume = {507},
       number = {1},
        pages = {1492-1512},
          doi = {10.1093/mnras/stab2318},
       adsurl = {https://ui.adsabs.harvard.edu/abs/2021MNRAS.507.1492B}
}

@ARTICLE{Stafford2019_radio_obs_SNR,
       author = {{Stafford}, Jennifer N. and {Lopez}, Laura A. and {Auchettl}, Katie and {Holland-Ashford}, Tyler},
        title = "{The Age Evolution of the Radio Morphology of Supernova Remnants}",
      journal = {\apj},
         year = 2019,
        month = oct,
       volume = {884},
       number = {2},
          eid = {113},
        pages = {113},
          doi = {10.3847/1538-4357/ab3a33},
archivePrefix = {arXiv},
       eprint = {1808.08234},
 primaryClass = {astro-ph.HE},
       adsurl = {https://ui.adsabs.harvard.edu/abs/2019ApJ...884..113S}
}

@article{Healpy2019,
  doi = {10.21105/joss.01298},
  url = {https://doi.org/10.21105/joss.01298},
  year = {2019},
  month = mar,
  publisher = {The Open Journal},
  volume = {4},
  number = {35},
  pages = {1298},
  author = {Andrea Zonca and Leo Singer and Daniel Lenz and Martin Reinecke and Cyrille Rosset and Eric Hivon and Krzysztof Gorski},
  title = {healpy: equal area pixelization and spherical harmonics transforms for data on the sphere in Python},
  journal = {Journal of Open Source Software}
}

@ARTICLE{HealPix2005,
   author = {{G{\'o}rski}, K.~M. and {Hivon}, E. and {Banday}, A.~J. and 
	{Wandelt}, B.~D. and {Hansen}, F.~K. and {Reinecke}, M. and 
	{Bartelmann}, M.},
    title = "{HEALPix: A Framework for High-Resolution Discretization and Fast Analysis of Data Distributed on the Sphere}",
  journal = {\apj},
   eprint = {arXiv:astro-ph/0409513},
     year = 2005,
    month = apr,
   volume = 622,
    pages = {759-771},
      doi = {10.1086/427976},
   adsurl = {http://adsabs.harvard.edu/abs/2005ApJ...622..759G}
}

@ARTICLE{Vishniac1983_overstability,
       author = {{Vishniac}, E.~T.},
        title = "{The dynamic and gravitational instabilities of spherical shocks}",
      journal = {\apj},
         year = 1983,
        month = nov,
       volume = {274},
        pages = {152-167},
          doi = {10.1086/161433},
       adsurl = {https://ui.adsabs.harvard.edu/abs/1983ApJ...274..152V}
}

@ARTICLE{Kim2015_SNR_expansion,
       author = {{Kim}, Chang-Goo and {Ostriker}, Eve C.},
        title = "{Momentum Injection by Supernovae in the Interstellar Medium}",
      journal = {\apj},
         year = 2015,
        month = apr,
       volume = {802},
       number = {2},
          eid = {99},
        pages = {99},
          doi = {10.1088/0004-637X/802/2/99},
archivePrefix = {arXiv},
       eprint = {1410.1537},
 primaryClass = {astro-ph.GA},
       adsurl = {https://ui.adsabs.harvard.edu/abs/2015ApJ...802...99K}
}

@ARTICLE{Walch2015_SNR,
       author = {{Walch}, Stefanie and {Naab}, Thorsten},
        title = "{The energy and momentum input of supernova explosions in structured and ionized molecular clouds}",
      journal = {\mnras},
         year = 2015,
        month = aug,
       volume = {451},
       number = {3},
        pages = {2757-2771},
          doi = {10.1093/mnras/stv1155},
archivePrefix = {arXiv},
       eprint = {1410.0011},
 primaryClass = {astro-ph.GA},
       adsurl = {https://ui.adsabs.harvard.edu/abs/2015MNRAS.451.2757W}
}

@misc{Jow2024cuspcuspsuniversalmodel,
      title={On the cusp of cusps: a universal model for extreme scattering events in the ISM}, 
      author={Dylan L. Jow and Ue-Li Pen and Daniel Baker},
      year={2024},
      eprint={2301.08344},
      archivePrefix={arXiv},
      primaryClass={astro-ph.HE},
      url={https://arxiv.org/abs/2301.08344}, 
}

@article{Ocker_2022_radio_scattering,
   title={Radio Scattering Horizons for Galactic and Extragalactic Transients},
   volume={934},
   ISSN={1538-4357},
   url={http://dx.doi.org/10.3847/1538-4357/ac75ba},
   DOI={10.3847/1538-4357/ac75ba},
   number={1},
   journal={The Astrophysical Journal},
   publisher={American Astronomical Society},
   author={Ocker, Stella Koch and Cordes, James M. and Chatterjee, Shami and Gorsuch, Miranda R.},
   year={2022},
   month=jul, pages={71} }

@misc{kempski2024unifiedmodelcosmicray,
      title={A Unified Model of Cosmic Ray Propagation and Radio Extreme Scattering Events from Intermittent Interstellar Structures}, 
      author={Philipp Kempski and Dongzi Li and Drummond B. Fielding and Eliot Quataert and E. Sterl Phinney and Matthew W. Kunz and Dylan L. Jow and Alexander A. Philippov},
      year={2024},
      eprint={2412.03649},
      archivePrefix={arXiv},
      primaryClass={astro-ph.HE},
      url={https://arxiv.org/abs/2412.03649}, 
}

@book{Dawson-Howe2014,
  author    = {Kenneth Dawson-Howe},
  title     = {A Practical Introduction to Computer Vision with OpenCV},
  publisher = {John Wiley \& Sons},
  year      = {2014},
  isbn13    = {9781118848456},
  isbn10    = {1118848454}
}

@ARTICLE{Lubke2025_CR_transport,
       author = {{L{\"u}bke}, Jeremiah and {Reichherzer}, Patrick and {Aerdker}, Sophie and {Effenberger}, Frederic and {Wilbert}, Mike and {Fichtner}, Horst and {Grauer}, Rainer},
        title = "{Modelling Cosmic-Ray Transport: Magnetized versus Unmagnetized Motion in Astrophysical Magnetic Turbulence}",
      journal = {arXiv e-prints},
         year = 2025,
        month = may,
          eid = {arXiv:2505.18155},
        pages = {arXiv:2505.18155},
          doi = {10.48550/arXiv.2505.18155},
archivePrefix = {arXiv},
       eprint = {2505.18155},
 primaryClass = {physics.plasm-ph},
       adsurl = {https://ui.adsabs.harvard.edu/abs/2025arXiv250518155L}
}

@ARTICLE{Kraichnan1967_two_dimensional_turbulence,
       author = {{Kraichnan}, Robert H.},
        title = "{Inertial Ranges in Two-Dimensional Turbulence}",
      journal = {Physics of Fluids},
         year = 1967,
        month = jul,
       volume = {10},
       number = {7},
        pages = {1417-1423},
          doi = {10.1063/1.1762301},
       adsurl = {https://ui.adsabs.harvard.edu/abs/1967PhFl...10.1417K}
}

@ARTICLE{Hennebelle2014_SN_driven_turb,
       author = {{Hennebelle}, Patrick and {Iffrig}, Olivier},
        title = "{Simulations of magnetized multiphase galactic disc regulated by supernovae explosions}",
      journal = {\aap},
         year = 2014,
        month = oct,
       volume = {570},
          eid = {A81},
        pages = {A81},
          doi = {10.1051/0004-6361/201423392},
archivePrefix = {arXiv},
       eprint = {1405.7819},
 primaryClass = {astro-ph.GA},
       adsurl = {https://ui.adsabs.harvard.edu/abs/2014A&A...570A..81H}
}

@ARTICLE{Guo2024_high_res_SN,
       author = {{Guo}, Minghao and {Kim}, Chang-Goo and {Stone}, James M.},
        title = "{Evolution of Supernova Remnants in a Cloudy Multiphase Interstellar Medium}",
      journal = {arXiv e-prints},
         year = 2024,
        month = nov,
          eid = {arXiv:2411.12809},
        pages = {arXiv:2411.12809},
          doi = {10.48550/arXiv.2411.12809},
archivePrefix = {arXiv},
       eprint = {2411.12809},
 primaryClass = {astro-ph.GA},
       adsurl = {https://ui.adsabs.harvard.edu/abs/2024arXiv241112809G}
}

@ARTICLE{Colman2022_large_scale_driving_in_ISM,
       author = {{Colman}, Tine and {Robitaille}, Jean-Fran{\c{c}}ois and {Hennebelle}, Patrick and {Miville-Desch{\^e}nes}, Marc-Antoine and {Brucy}, No{\'e} and {Klessen}, Ralf S. and {Glover}, Simon C.~O. and {Soler}, Juan D. and {Elia}, Davide and {Traficante}, Alessio and {Molinari}, Sergio and {Testi}, Leonardo},
        title = "{The signature of large-scale turbulence driving on the structure of the interstellar medium}",
      journal = {The Monthly Notices of The Royal Astronomical Society},
         year = 2022,
        month = aug,
       volume = {514},
       number = {3},
        pages = {3670-3684},
          doi = {10.1093/mnras/stac1543},
archivePrefix = {arXiv},
       eprint = {2206.00451},
 primaryClass = {astro-ph.GA},
       adsurl = {https://ui.adsabs.harvard.edu/abs/2022MNRAS.514.3670C}
}

@ARTICLE{Kapyla2018_vorticity_helicity_ISM,
       author = {{K{\"a}pyl{\"a}}, M.~J. and {Gent}, F.~A. and {V{\"a}is{\"a}l{\"a}}, M.~S. and {Sarson}, G.~R.},
        title = "{The supernova-regulated ISM. III. Generation of vorticity, helicity, and mean flows}",
      journal = {\aap},
         year = 2018,
        month = mar,
       volume = {611},
          eid = {A15},
        pages = {A15},
          doi = {10.1051/0004-6361/201731228},
archivePrefix = {arXiv},
       eprint = {1705.08642},
 primaryClass = {astro-ph.GA},
       adsurl = {https://ui.adsabs.harvard.edu/abs/2018A&A...611A..15K}
}

@ARTICLE{Malvadi2023_numerical_diss,
       author = {{Shivakumar}, Lakshmi Malvadi and {Federrath}, Christoph},
        title = "{Numerical viscosity and resistivity in MHD turbulence simulations}",
      journal = {\mnras},
         year = 2025,
        month = mar,
       volume = {537},
       number = {4},
        pages = {2961-2986},
          doi = {10.1093/mnras/staf160},
archivePrefix = {arXiv},
       eprint = {2311.10350},
 primaryClass = {astro-ph.SR},
       adsurl = {https://ui.adsabs.harvard.edu/abs/2025MNRAS.537.2961S}
}

@article{scikit-image,
 title = {scikit-image: image processing in {P}ython},
 author = {van der Walt, {S}t\'efan and {S}ch\"onberger, {J}ohannes {L}. and
           {Nunez-Iglesias}, {J}uan and {B}oulogne, {F}ran\c{c}ois and {W}arner,
           {J}oshua {D}. and {Y}ager, {N}eil and {G}ouillart, {E}mmanuelle and
           {Y}u, {T}ony and the scikit-image contributors},
 year = {2014},
 month = {6},
 volume = {2},
 pages = {453},
 journal = {PeerJ},
 issn = {2167-8359},
 url = {http://dx.doi.org/10.7717/peerj.453},
 doi = {10.7717/peerj.453}
}

@article{MacLow2004,
archivePrefix = {arXiv},
arxivId = {astro-ph/0301093},
author = {Mac Low, Mordecai Mark and Klessen, Ralf S.},
doi = {10.1103/RevModPhys.76.125},
eprint = {0301093},
isbn = {0034-6861},
issn = {00346861},
journal = {Reviews of Modern Physics},
number = {1},
pages = {125--194},
primaryClass = {astro-ph},
title = {{Control of star formation by supersonic turbulence}},
volume = {76},
year = {2004}
}

@article{Federrath2010_solendoidal_versus_compressive,
archivePrefix = {arXiv},
arxivId = {0905.1060},
author = {Federrath, C. and Roman-Duval, J. and Klessen, R. and Schmidt, W. and {Mac Low}, M. -M.},
doi = {10.1051/0004-6361/200912437},
eprint = {0905.1060},
issn = {0004-6361},
journal = {Astronomy and Astrophysics},
number = {A81},
title = {{Comparing the statistics of interstellar turbulence in simulations and observations: Solenoidal versus compressive turbulence forcing}},
url = {http://arxiv.org/abs/0905.1060{\%}0Ahttp://dx.doi.org/10.1051/0004-6361/200912437},
volume = {512},
year = {2010}
}

@article{Elmegreen2004,
archivePrefix = {arXiv},
arxivId = {astro-ph/0404451},
author = {Elmegreen, Bruce G. and Scalo, John},
doi = {10.1146/annurev.astro.41.011802.094859},
eprint = {0404451},
isbn = {00664146 (ISSN)},
issn = {0066-4146},
journal = {Annu. Rev. Astron. Astrophys.},
pages = {211--273},
primaryClass = {astro-ph},
title = {{Interstellar Turbulence I: Observations and Processes}},
url = {http://arxiv.org/abs/astro-ph/0404451{\%}0Ahttp://dx.doi.org/10.1146/annurev.astro.41.011802.094859},
volume = {42},
year = {2004}
}

@article{Federrath2016_filaments,
archivePrefix = {arXiv},
arxivId = {1510.05654},
author = {Federrath, Christoph},
doi = {10.1093/mnras/stv2880},
eprint = {1510.05654},
issn = {13652966},
journal = {The Monthly Notices of The Royal Astronomical Society},
number = {1},
pages = {375--388},
title = {{On the universality of interstellar filaments: Theory meets simulations and observations}},
volume = {457},
year = {2016}
}

@ARTICLE{Ferriere2020_reynolds_numbers_for_ism,
       author = {{Ferri{\`e}re}, K.},
        title = "{Plasma turbulence in the interstellar medium}",
      journal = {Plasma Physics and Controlled Fusion},
         year = 2020,
        month = jan,
       volume = {62},
       number = {1},
        pages = {014014},
          doi = {10.1088/1361-6587/ab49eb},
archivePrefix = {arXiv},
       eprint = {1912.08237},
 primaryClass = {astro-ph.GA},
       adsurl = {https://ui.adsabs.harvard.edu/abs/2020PPCF...62a4014F}
}

@ARTICLE{Krumholz2018_metallicity_SF,
       author = {{Krumholz}, Mark R. and {Ting}, Yuan-Sen},
        title = "{Metallicity fluctuation statistics in the interstellar medium and young stars - I. Variance and correlation}",
      journal = {The Monthly Notices of The Royal Astronomical Society},
         year = 2018,
        month = apr,
       volume = {475},
       number = {2},
        pages = {2236-2252},
          doi = {10.1093/mnras/stx3286},
archivePrefix = {arXiv},
       eprint = {1708.06853},
 primaryClass = {astro-ph.GA},
       adsurl = {https://ui.adsabs.harvard.edu/abs/2018MNRAS.475.2236K}
}

@article{Boldyrev2006,
  title = {Spectrum of Magnetohydrodynamic Turbulence},
  author = {Boldyrev, Stanislav},
  journal = {Physical Review Letters},
  volume = {96},
  issue = {11},
  pages = {115002},
  numpages = {4},
  year = {2006},
  month = {Mar},
  publisher = {American Physical Society},
  doi = {10.1103/PhysRevLett.96.115002},
  url = {https://link.aps.org/doi/10.1103/PhysRevLett.96.115002}
}

@ARTICLE{Bacchini2020_supernova_drives_turb,
       author = {{Bacchini}, Cecilia and {Fraternali}, Filippo and {Iorio}, Giuliano and {Pezzulli}, Gabriele and {Marasco}, Antonino and {Nipoti}, Carlo},
        title = "{Evidence for supernova feedback sustaining gas turbulence in nearby star-forming galaxies}",
      journal = {\aap},
         year = 2020,
        month = sep,
       volume = {641},
          eid = {A70},
        pages = {A70},
          doi = {10.1051/0004-6361/202038223},
archivePrefix = {arXiv},
       eprint = {2006.10764},
 primaryClass = {astro-ph.GA},
       adsurl = {https://ui.adsabs.harvard.edu/abs/2020A&A...641A..70B}
}

@INPROCEEDINGS{numba:2015,
       author = {{Lam}, Siu Kwan and {Pitrou}, Antoine and {Seibert}, Stanley},
        title = "{Numba: A LLVM-based Python JIT Compiler}",
    booktitle = {Proc. Second Workshop on the LLVM Compiler Infrastructure in HPC},
         year = 2015,
        month = nov,
        pages = {1-6},
          doi = {10.1145/2833157.2833162},
       adsurl = {https://ui.adsabs.harvard.edu/abs/2015llvm.confE...1L}
}

@ARTICLE{Fielding2018_winds,
       author = {{Fielding}, Drummond and {Quataert}, Eliot and {Martizzi}, Davide},
        title = "{Clustered supernovae drive powerful galactic winds after superbubble breakout}",
      journal = {The Monthly Notices of The Royal Astronomical Society},
         year = 2018,
        month = dec,
       volume = {481},
       number = {3},
        pages = {3325-3347},
          doi = {10.1093/mnras/sty2466},
archivePrefix = {arXiv},
       eprint = {1807.08758},
 primaryClass = {astro-ph.GA},
       adsurl = {https://ui.adsabs.harvard.edu/abs/2018MNRAS.481.3325F}
}

@ARTICLE{Kolborg2023_metal_mixing_2,
       author = {{Kolborg}, Anne Noer and {Ramirez-Ruiz}, Enrico and {Martizzi}, Davide and {Macias}, Phillip and {Soares-Furtado}, Melinda},
        title = "{Constraints on the Frequency and Mass Content of r-process Events Derived from Turbulent Mixing in Galactic Disks}",
      journal = {The Astrophysical Journal},
         year = 2023,
        month = jun,
       volume = {949},
       number = {2},
          eid = {100},
        pages = {100},
          doi = {10.3847/1538-4357/acca80},
archivePrefix = {arXiv},
       eprint = {2304.01144},
 primaryClass = {astro-ph.GA},
       adsurl = {https://ui.adsabs.harvard.edu/abs/2023ApJ...949..100K}
}

@ARTICLE{Kolborg2022_metal_mixing_1,
       author = {{Kolborg}, Anne Noer and {Martizzi}, Davide and {Ramirez-Ruiz}, Enrico and {Pfister}, Hugo and {Sakari}, Charli and {Wechsler}, Risa H. and {Soares-Furtado}, Melinda},
        title = "{Supernova-driven Turbulent Metal Mixing in High-redshift Galactic Disks: Metallicity Fluctuations in the Interstellar Medium and its Imprints on Metal-poor Stars in the Milky Way}",
      journal = {The Astrophysical Journal Letters},
         year = 2022,
        month = sep,
       volume = {936},
       number = {2},
          eid = {L26},
        pages = {L26},
          doi = {10.3847/2041-8213/ac8c98},
archivePrefix = {arXiv},
       eprint = {2111.02619},
 primaryClass = {astro-ph.GA},
       adsurl = {https://ui.adsabs.harvard.edu/abs/2022ApJ...936L..26K}
}

@ARTICLE{Beattie2025_nature_astro,
       author = {{Beattie}, James R. and {Federrath}, Christoph and {Klessen}, Ralf S. and {Cielo}, Salvatore and {Bhattacharjee}, Amitava},
        title = "{The spectrum of magnetized turbulence in the interstellar medium}",
      journal = {Nature Astronomy},
         year = 2025,
        month = may,
          doi = {10.1038/s41550-025-02551-5},
archivePrefix = {arXiv},
       eprint = {2504.07136},
 primaryClass = {astro-ph.GA},
       adsurl = {https://ui.adsabs.harvard.edu/abs/2025NatAs.tmp..114B}
}

@ARTICLE{Kriel2025_SSD,
       author = {{Kriel}, Neco and {Beattie}, James R. and {Federrath}, Christoph and {Krumholz}, Mark R. and {Hew}, Justin Kin Jun},
        title = "{Fundamental MHD scales - II. The kinematic phase of the supersonic small-scale dynamo}",
      journal = {\mnras},
         year = 2025,
        month = mar,
       volume = {537},
       number = {3},
        pages = {2602-2629},
          doi = {10.1093/mnras/staf188},
archivePrefix = {arXiv},
       eprint = {2310.17036},
 primaryClass = {astro-ph.GA},
       adsurl = {https://ui.adsabs.harvard.edu/abs/2025MNRAS.537.2602K}
}

@ARTICLE{Kriel2025_nonlinear_dynamo,
       author = {{Kriel}, Neco and {Beattie}, James R. and {Krumholz}, Mark R. and {Schober}, Jennifer and {Armstrong}, Patrick J.},
        title = "{The growth of magnetic energy during the nonlinear phase of the subsonic and supersonic small-scale dynamo}",
      journal = {arXiv e-prints},
         year = 2025,
        month = sep,
          eid = {arXiv:2509.09949},
        pages = {arXiv:2509.09949},
          doi = {10.48550/arXiv.2509.09949},
archivePrefix = {arXiv},
       eprint = {2509.09949},
 primaryClass = {physics.plasm-ph},
       adsurl = {https://ui.adsabs.harvard.edu/abs/2025arXiv250909949K}
}

@ARTICLE{Sampson2025_CRMHD_heating,
       author = {{Sampson}, Matt L. and {Beattie}, James R. and {Teyssier}, Romain and {Kempski}, Philipp and {Moseley}, Eric R. and {Commer{\c{c}}on}, Beno{\^\i}t and {Dubois}, Yohan and {Rosdahl}, Joakim},
        title = "{Cosmic ray and plasma coupling for isothermal supersonic turbulence in the magnetized interstellar medium}",
      journal = {arXiv e-prints},
         year = 2025,
        month = jun,
          eid = {arXiv:2506.03768},
        pages = {arXiv:2506.03768},
          doi = {10.48550/arXiv.2506.03768},
archivePrefix = {arXiv},
       eprint = {2506.03768},
 primaryClass = {astro-ph.GA},
       adsurl = {https://ui.adsabs.harvard.edu/abs/2025arXiv250603768S}
}

@ARTICLE{Velden2020_cmasher,
       author = {{van der Velden}, Ellert},
        title = "{CMasher: Scientific colormaps for making accessible, informative and 'cmashing' plots}",
      journal = {The Journal of Open Source Software},
         year = 2020,
        month = feb,
       volume = {5},
       number = {46},
          eid = {2004},
        pages = {2004},
          doi = {10.21105/joss.02004},
archivePrefix = {arXiv},
       eprint = {2003.01069},
 primaryClass = {eess.IV},
       adsurl = {https://ui.adsabs.harvard.edu/abs/2020JOSS....5.2004V}
}

@ARTICLE{Iroshnikov_1965_IK_turb,
       author = {{Iroshnikov}, P.~S.},
        title = "{Turbulence of a Conducting Fluid in a Strong Magnetic Field}",
      journal = {Soviet Astronomy},
         year = 1964,
        month = feb,
       volume = {7},
        pages = {566},
       adsurl = {https://ui.adsabs.harvard.edu/abs/1964SvA.....7..566I}
}

@ARTICLE{Kriel2022_kinematic_dynamo_scales,
       author = {{Kriel}, Neco and {Beattie}, James R. and {Seta}, Amit and {Federrath}, Christoph},
        title = "{Fundamental scales in the kinematic phase of the turbulent dynamo}",
      journal = {The Monthly Notices of The Royal Astronomical Society},
         year = 2022,
        month = jun,
       volume = {513},
       number = {2},
        pages = {2457-2470},
          doi = {10.1093/mnras/stac969},
archivePrefix = {arXiv},
       eprint = {2204.00828},
 primaryClass = {astro-ph.SR},
       adsurl = {https://ui.adsabs.harvard.edu/abs/2022MNRAS.513.2457K}
}

@ARTICLE{Beattie2025_bulk_viscosity,
       author = {{Beattie}, James R. and {Federrath}, Christoph and {Kriel}, Neco and {Hew}, Justin Kin Jun and {Bhattacharjee}, Amitava},
        title = "{Taking control of compressible modes: bulk viscosity and the turbulent dynamo}",
      journal = {\mnras},
         year = 2025,
        month = aug,
          doi = {10.1093/mnras/staf1318},
archivePrefix = {arXiv},
       eprint = {2312.03984},
 primaryClass = {astro-ph.GA},
       adsurl = {https://ui.adsabs.harvard.edu/abs/2025MNRAS.tmp.1274B}
}

@ARTICLE{Balsara2004_SNe_driven_turb,
       author = {{Balsara}, Dinshaw S. and {Kim}, Jongsoo and {Mac Low}, Mordecai-Mark and {Mathews}, Grant J.},
        title = "{Amplification of Interstellar Magnetic Fields by Supernova-driven Turbulence}",
      journal = {\apj},
         year = 2004,
        month = dec,
       volume = {617},
       number = {1},
        pages = {339-349},
          doi = {10.1086/425297},
archivePrefix = {arXiv},
       eprint = {astro-ph/0403660},
 primaryClass = {astro-ph},
       adsurl = {https://ui.adsabs.harvard.edu/abs/2004ApJ...617..339B}
}

@ARTICLE{Kochurin2024_weak_versus_strong_turb,
       author = {{Kochurin}, E.~A. and {Kuznetsov}, E.~A.},
        title = "{Three-Dimensional Acoustic Turbulence: Weak Versus Strong}",
      journal = {\prl},
         year = 2024,
        month = nov,
       volume = {133},
       number = {20},
          eid = {207201},
        pages = {207201},
          doi = {10.1103/PhysRevLett.133.207201},
archivePrefix = {arXiv},
       eprint = {2407.08352},
 primaryClass = {nlin.CD},
       adsurl = {https://ui.adsabs.harvard.edu/abs/2024PhRvL.133t7201K}
}

@ARTICLE{Hill2012_SNe_driven_turb,
       author = {{Hill}, Alex S. and {Joung}, M. Ryan and {Mac Low}, Mordecai-Mark and {Benjamin}, Robert A. and {Haffner}, L. Matthew and {Klingenberg}, Christian and {Waagan}, Knut},
        title = "{Vertical Structure of a Supernova-driven Turbulent, Magnetized Interstellar Medium}",
      journal = {\apj},
         year = 2012,
        month = may,
       volume = {750},
       number = {2},
          eid = {104},
        pages = {104},
          doi = {10.1088/0004-637X/750/2/104},
archivePrefix = {arXiv},
       eprint = {1202.0552},
 primaryClass = {astro-ph.GA},
       adsurl = {https://ui.adsabs.harvard.edu/abs/2012ApJ...750..104H}
}

@ARTICLE{Beattie2025_sda,
       author = {{Beattie}, James R. and {Bhattacharjee}, Amitava},
        title = "{Scale-dependent alignment in compressible magnetohydrodynamic turbulence}",
      journal = {arXiv e-prints},
         year = 2025,
        month = apr,
          eid = {arXiv:2504.15538},
        pages = {arXiv:2504.15538},
          doi = {10.48550/arXiv.2504.15538},
archivePrefix = {arXiv},
       eprint = {2504.15538},
 primaryClass = {physics.plasm-ph},
       adsurl = {https://ui.adsabs.harvard.edu/abs/2025arXiv250415538B}
}

@ARTICLE{Stone2024_athenaK,
       author = {{Stone}, James M. and {Mullen}, Patrick D. and {Fielding}, Drummond and {Grete}, Philipp and {Guo}, Minghao and {Kempski}, Philipp and {Most}, Elias R. and {White}, Christopher J. and {Wong}, George N.},
        title = "{AthenaK: A Performance-Portable Version of the Athena++ AMR Framework}",
      journal = {arXiv e-prints},
         year = 2024,
        month = sep,
          eid = {arXiv:2409.16053},
        pages = {arXiv:2409.16053},
          doi = {10.48550/arXiv.2409.16053},
archivePrefix = {arXiv},
       eprint = {2409.16053},
 primaryClass = {astro-ph.IM},
       adsurl = {https://ui.adsabs.harvard.edu/abs/2024arXiv240916053S}
}

@software{2021ascl.soft09009G,
       author = {{Gomersall}, Henry},
        title = "{pyFFTW: Python wrapper around FFTW}",
 howpublished = {Astrophysics Source Code Library, record ascl:2109.009},
         year = 2021,
        month = sep,
          eid = {ascl:2109.009},
       adsurl = {https://ui.adsabs.harvard.edu/abs/2021ascl.soft09009G}
}

@ARTICLE{Galtier2023_fast_wave_turb,
       author = {{Galtier}, S{\'e}bastien},
        title = "{Fast magneto-acoustic wave turbulence and the Iroshnikov-Kraichnan spectrum}",
      journal = {Journal of Plasma Physics},
         year = 2023,
        month = apr,
       volume = {89},
       number = {2},
          eid = {905890205},
        pages = {905890205},
          doi = {10.1017/S0022377823000259},
archivePrefix = {arXiv},
       eprint = {2303.00643},
 primaryClass = {physics.plasm-ph},
       adsurl = {https://ui.adsabs.harvard.edu/abs/2023JPlPh..89b9005G}
}

@ARTICLE{Mee2006_blastwave_turbulence,
       author = {{Mee}, Antony J. and {Brandenburg}, Axel},
        title = "{Turbulence from localized random expansion waves}",
      journal = {\mnras},
         year = 2006,
        month = jul,
       volume = {370},
       number = {1},
        pages = {415-419},
          doi = {10.1111/j.1365-2966.2006.10476.x},
archivePrefix = {arXiv},
       eprint = {astro-ph/0602057},
 primaryClass = {astro-ph},
       adsurl = {https://ui.adsabs.harvard.edu/abs/2006MNRAS.370..415M}
}

@ARTICLE{McLeod2013_simulations_of_the_nonlinearVish,
       author = {{McLeod}, A.~D. and {Whitworth}, A.~P.},
        title = "{Simulations of the non-linear thin shell instability}",
      journal = {\mnras},
         year = 2013,
        month = may,
       volume = {431},
       number = {1},
        pages = {710-721},
          doi = {10.1093/mnras/stt203},
archivePrefix = {arXiv},
       eprint = {1302.1867},
 primaryClass = {astro-ph.SR},
       adsurl = {https://ui.adsabs.harvard.edu/abs/2013MNRAS.431..710M}
}

@ARTICLE{Connor2025_cascading_from_the_winds,
       author = {{Connor}, Isabelle and {Beattie}, James R. and {Noer Kolborg}, Anne and {Ramirez-Ruiz}, Enrico},
        title = "{Cascading from the winds to the disk: the universality of supernovae-driven turbulence in different galactic interstellar mediums}",
      journal = {arXiv e-prints},
         year = 2025,
        month = sep,
          eid = {arXiv:2509.01653},
        pages = {arXiv:2509.01653},
archivePrefix = {arXiv},
       eprint = {2509.01653},
 primaryClass = {astro-ph.GA},
       adsurl = {https://ui.adsabs.harvard.edu/abs/2025arXiv250901653C}
}

@Article{Hunter2007,
  Author    = {Hunter, J. D.},
  Title     = {Matplotlib: A 2D graphics environment},
  Journal   = {Computing in Science \& Engineering},
  Volume    = {9},
  Number    = {3},
  Pages     = {90--95},
  publisher = {IEEE COMPUTER SOC},
  doi       = {10.1109/MCSE.2007.55},
  year      = 2007
}

@ARTICLE{Virtanen2020,
       author = {{Virtanen}, Pauli and {Gommers}, Ralf and {Oliphant},
         Travis E. and {Haberland}, Matt and {Reddy}, Tyler and
         {Cournapeau}, David and {Burovski}, Evgeni and {Peterson}, Pearu
         and {Weckesser}, Warren and {Bright}, Jonathan and {van der Walt},
         St{\'e}fan J.  and {Brett}, Matthew and {Wilson}, Joshua and
         {Jarrod Millman}, K.  and {Mayorov}, Nikolay and {Nelson}, Andrew
         R.~J. and {Jones}, Eric and {Kern}, Robert and {Larson}, Eric and
         {Carey}, CJ and {Polat}, {\.I}lhan and {Feng}, Yu and {Moore},
         Eric W. and {Vand erPlas}, Jake and {Laxalde}, Denis and
         {Perktold}, Josef and {Cimrman}, Robert and {Henriksen}, Ian and
         {Quintero}, E.~A. and {Harris}, Charles R and {Archibald}, Anne M.
         and {Ribeiro}, Ant{\^o}nio H. and {Pedregosa}, Fabian and
         {van Mulbregt}, Paul and {Contributors}, SciPy 1. 0},
        title = "{SciPy 1.0: Fundamental Algorithms for Scientific
                  Computing in Python}",
      journal = {Nature Methods},
      year = "2020",
      volume={17},
      pages={261--272},
      adsurl = {https://rdcu.be/b08Wh},
      doi = {https://doi.org/10.1038/s41592-019-0686-2},
}

@Misc{Oliphant2006,
  author =    {Travis Oliphant},
  title =     {{NumPy}: A guide to {NumPy}},
  year =      {2006},
  howpublished = {USA: Trelgol Publishing},
  url = "http://www.numpy.org/",
  note = {[Online; accessed <today>]}
 }

@InCollection{Childs2012,
    author = {Hank Childs and Eric Brugger and Brad Whitlock and Jeremy Meredith and Sean Ahern and David Pugmire and Kathleen Biagas and Mark Miller and Cyrus Harrison and Gunther H. Weber and Hari Krishnan and Thomas Fogal and Allen Sanderson and Christoph Garth and E. Wes Bethel and David Camp and Oliver R\"{u}bel and Marc Durant and Jean M. Favre and Paul Navr\'{a}til},
    title = {{VisIt: An End-User Tool For Visualizing and Analyzing Very Large Data}},
    year = "2012",
    pages = "357-372",
    month = "Oct",
    booktitle = {{High Performance Visualization--Enabling Extreme-Scale Scientific Insight}},
    publisher = {Taylor \& Francis}
}

@ARTICLE{Beuther2015,
       author = {{Beuther}, H. and {Ragan}, S.~E. and {Johnston}, K. and {Henning}, Th. and
         {Hacar}, A. and {Kainulainen}, J.~T.},
        title = "{Filament fragmentation in high-mass star formation}",
      journal = {Astronomy and Astrophysics},
         year = 2015,
        month = dec,
       volume = {584},
          eid = {A67},
        pages = {A67},
          doi = {10.1051/0004-6361/201527108},
archivePrefix = {arXiv},
       eprint = {1510.07063},
 primaryClass = {astro-ph.GA},
       adsurl = {https://ui.adsabs.harvard.edu/abs/2015A&A...584A..67B}
}

@article{Behnel2011, 
  title={Cython: The best of both worlds}, 
  author={Behnel, Stefan and Bradshaw, Robert and Citro, Craig and Dalcin, Lisandro and Seljebotn, Dag Sverre and Smith, Kurt}, 
  journal={Computing in Science \& Engineering}, 
  volume={13}, 
  number={2}, 
  pages={31--39}, 
  year={2011}, 
  publisher={IEEE} 
}

@ARTICLE{Federrath2014_supersonic_dynamo,
       author = {{Federrath}, Christoph and {Schober}, Jennifer and {Bovino}, Stefano and
         {Schleicher}, Dominik R.~G.},
        title = "{The Turbulent Dynamo in Highly Compressible Supersonic Plasmas}",
      journal = {The Astrophysical Journal Letters},
         year = 2014,
        month = dec,
       volume = {797},
       number = {2},
          eid = {L19},
        pages = {L19},
          doi = {10.1088/2041-8205/797/2/L19},
archivePrefix = {arXiv},
       eprint = {1411.4707},
 primaryClass = {astro-ph.GA},
       adsurl = {https://ui.adsabs.harvard.edu/abs/2014ApJ...797L..19F}
}

@ARTICLE{Kazantsev1968,
       author = {{Kazantsev}, A.~P.},
        title = "{Enhancement of a Magnetic Field by a Conducting Fluid}",
      journal = {Soviet Journal of Experimental and Theoretical Physics},
         year = 1968,
        month = may,
       volume = {26},
        pages = {1031},
       adsurl = {https://ui.adsabs.harvard.edu/abs/1968JETP...26.1031K}
}

@ARTICLE{Padoan2016_supernova_driving,
       author = {{Padoan}, Paolo and {Pan}, Liubin and {Haugb{\o}lle}, Troels and {Nordlund}, {\r{A}}ke},
        title = "{Supernova Driving. I. The Origin of Molecular Cloud Turbulence}",
      journal = {The Astrophysical Journal},
         year = 2016,
        month = may,
       volume = {822},
       number = {1},
          eid = {11},
        pages = {11},
          doi = {10.3847/0004-637X/822/1/11},
archivePrefix = {arXiv},
       eprint = {1509.04663},
 primaryClass = {astro-ph.GA},
       adsurl = {https://ui.adsabs.harvard.edu/abs/2016ApJ...822...11P}
}

@ARTICLE{Robertson2018,
       author = {{Robertson}, Brant and {Goldreich}, Peter},
        title = "{Dense Regions in Supersonic Isothermal Turbulence}",
      journal = {The Astrophysical Journal},
         year = 2018,
        month = feb,
       volume = {854},
       number = {2},
          eid = {88},
        pages = {88},
          doi = {10.3847/1538-4357/aaa89e},
archivePrefix = {arXiv},
       eprint = {1801.05440},
 primaryClass = {astro-ph.GA},
       adsurl = {https://ui.adsabs.harvard.edu/abs/2018ApJ...854...88R}
}

@article{vanderWalts2014,
 title = {scikit-image: image processing in {P}ython},
 author = {van der Walt, {S}t\'efan and {S}ch\"onberger, {J}ohannes {L}. and
           {Nunez-Iglesias}, {J}uan and {B}oulogne, {F}ran\c{c}ois and {W}arner,
           {J}oshua {D}. and {Y}ager, {N}eil and {G}ouillart, {E}mmanuelle and
           {Y}u, {T}ony and the scikit-image contributors},
 year = {2014},
 month = {6},
 volume = {2},
 pages = {e453},
 journal = {PeerJ},
 issn = {2167-8359},
 url = {https://doi.org/10.7717/peerj.453},
 doi = {10.7717/peerj.453}
}

@book{Stroustrup2013, 
author = {Stroustrup, Bjarne}, 
title = {The C++ Programming Language}, 
year = {2013}, 
isbn = {0321563840}, 
publisher = {Addison-Wesley Professional}, 
edition = {4th} 
}

@article{numpy2020,
	Author = {Harris, Charles R. and Millman, K. Jarrod and van der Walt, St{\'e}fan J. and Gommers, Ralf and Virtanen, Pauli and Cournapeau, David and Wieser, Eric and Taylor, Julian and Berg, Sebastian and Smith, Nathaniel J. and Kern, Robert and Picus, Matti and Hoyer, Stephan and van Kerkwijk, Marten H. and Brett, Matthew and Haldane, Allan and del R{\'\i}o, Jaime Fern{\'a}ndez and Wiebe, Mark and Peterson, Pearu and G{\'e}rard-Marchant, Pierre and Sheppard, Kevin and Reddy, Tyler and Weckesser, Warren and Abbasi, Hameer and Gohlke, Christoph and Oliphant, Travis E.},
	Da = {2020/09/01},
	Doi = {10.1038/s41586-020-2649-2},
	Id = {Harris2020},
	Isbn = {1476-4687},
	Journal = {Nature},
	Number = {7825},
	Pages = {357--362},
	Title = {Array programming with NumPy},
	Ty = {JOUR},
	Url = {https://doi.org/10.1038/s41586-020-2649-2},
	Volume = {585},
	Year = {2020}
}

@ARTICLE{Falgarone1995,
       author = {{Falgarone}, E. and {Pineau des Forets}, G. and {Roueff}, E.},
        title = "{Chemical signatures of the intermittency of turbulence in low density interstellar clouds.}",
      journal = {Astronomy and Astrophysics},
         year = 1995,
        month = aug,
       volume = {300},
        pages = {870},
       adsurl = {https://ui.adsabs.harvard.edu/abs/1995A&A...300..870F}
}

@ARTICLE{Falgarone2009,
       author = {{Falgarone}, E. and {Pety}, J. and {Hily-Blant}, P.},
        title = "{Intermittency of interstellar turbulence: extreme velocity-shears and CO emission on milliparsec scale}",
      journal = {Astronomy and Astrophysics},
         year = 2009,
        month = nov,
       volume = {507},
       number = {1},
        pages = {355-368},
          doi = {10.1051/0004-6361/200810963},
archivePrefix = {arXiv},
       eprint = {0910.1766},
 primaryClass = {astro-ph.GA},
       adsurl = {https://ui.adsabs.harvard.edu/abs/2009A&A...507..355F}
}

@ARTICLE{Hopkins2013_fragmentation,
       author = {{Hopkins}, Philip F.},
        title = "{A general theory of turbulent fragmentation}",
      journal = {The Monthly Notices of The Royal Astronomical Society},
         year = 2013,
        month = apr,
       volume = {430},
       number = {3},
        pages = {1653-1693},
          doi = {10.1093/mnras/sts704},
archivePrefix = {arXiv},
       eprint = {1210.0903},
 primaryClass = {astro-ph.CO},
       adsurl = {https://ui.adsabs.harvard.edu/abs/2013MNRAS.430.1653H}
}

@article{Federrath2021,
	Author = {Federrath, Christoph and Klessen, Ralf S. and Iapichino, Luigi and Beattie, James R.},
	Da = {2021/01/11},
	Doi = {10.1038/s41550-020-01282-z},
	Id = {Federrath2021},
	Isbn = {2397-3366},
	Journal = {Nature Astronomy},
	Title = {The sonic scale of interstellar turbulence},
	Ty = {JOUR},
	Url = {https://doi.org/10.1038/s41550-020-01282-z},
	Year = {2021}}

@ARTICLE{Ruszkowski2023_CRs_in_galaxies_review,
       author = {{Ruszkowski}, Mateusz and {Pfrommer}, Christoph},
        title = "{Cosmic ray feedback in galaxies and galaxy clusters -- A pedagogical introduction and a topical review of the acceleration, transport, observables, and dynamical impact of cosmic rays}",
      journal = {arXiv e-prints},
         year = 2023,
        month = jun,
          eid = {arXiv:2306.03141},
        pages = {arXiv:2306.03141},
          doi = {10.48550/arXiv.2306.03141},
archivePrefix = {arXiv},
       eprint = {2306.03141},
 primaryClass = {astro-ph.HE},
       adsurl = {https://ui.adsabs.harvard.edu/abs/2023arXiv230603141R}
}

@ARTICLE{Galishnikova2022_saturation_and_tearing,
       author = {{Galishnikova}, Alisa K. and {Kunz}, Matthew W. and {Schekochihin}, Alexander A.},
        title = "{Tearing Instability and Current-Sheet Disruption in the Turbulent Dynamo}",
      journal = {Physical Review X},
         year = 2022,
        month = oct,
       volume = {12},
       number = {4},
          eid = {041027},
        pages = {041027},
          doi = {10.1103/PhysRevX.12.041027},
archivePrefix = {arXiv},
       eprint = {2201.07757},
 primaryClass = {astro-ph.HE},
       adsurl = {https://ui.adsabs.harvard.edu/abs/2022PhRvX..12d1027G}
}

@ARTICLE{Gent2022_multiphase_dynamo,
       author = {{Gent}, Frederick A. and {Mac Low}, Mordecai-Mark and {Korpi-Lagg}, Maarit J. and {Singh}, Nishant K.},
        title = "{The Small-scale Dynamo in a Multiphase Supernova-driven Medium}",
      journal = {The Astrophysical Journal},
         year = 2023,
        month = feb,
       volume = {943},
       number = {2},
          eid = {176},
        pages = {176},
          doi = {10.3847/1538-4357/acac20},
archivePrefix = {arXiv},
       eprint = {2210.04460},
 primaryClass = {astro-ph.GA},
       adsurl = {https://ui.adsabs.harvard.edu/abs/2023ApJ...943..176G}
}

@ARTICLE{Seta2022_multiphase_dynamo,
       author = {{Seta}, Amit and {Federrath}, Christoph},
        title = "{Turbulent dynamo in the two-phase interstellar medium}",
      journal = {The Monthly Notices of The Royal Astronomical Society},
         year = 2022,
        month = jul,
       volume = {514},
       number = {1},
        pages = {957-976},
          doi = {10.1093/mnras/stac1400},
archivePrefix = {arXiv},
       eprint = {2202.08324},
 primaryClass = {astro-ph.GA},
       adsurl = {https://ui.adsabs.harvard.edu/abs/2022MNRAS.514..957S}
}

@ARTICLE{Armstrong1995_power_law,
       author = {{Armstrong}, J.~W. and {Rickett}, B.~J. and {Spangler}, S.~R.},
        title = "{Electron Density Power Spectrum in the Local Interstellar Medium}",
      journal = {The Astrophysical Journal},
         year = 1995,
        month = apr,
       volume = {443},
        pages = {209},
          doi = {10.1086/175515},
       adsurl = {https://ui.adsabs.harvard.edu/abs/1995ApJ...443..209A}
}

@article{Schekochihin2004_dynamo,
	doi = {10.1086/422547},
	url = {https://doi.org/10.1086/422547},
	year = 2004,
	month = {sep},
	publisher = {American Astronomical Society},
	volume = {612},
	number = {1},
	pages = {276--307},
	author = {Alexander A. Schekochihin and Steven C. Cowley and Samuel F. Taylor and Jason L. Maron and James C. McWilliams},
	title = {Simulations of the Small-Scale Turbulent Dynamo},
	journal = {The Astrophysical Journal}
}

@article{Kraichnan1965_IKturb,
    author = {Kraichnan,Robert H.},
    title = {Inertial‐Range Spectrum of Hydromagnetic Turbulence},
    journal = {The Physics of Fluids},
    volume = {8},
    number = {7},
    pages = {1385-1387},
    year = {1965},
    doi = {10.1063/1.1761412},
    URL = {https://aip.scitation.org/doi/abs/10.1063/1.1761412},
    eprint = {https://aip.scitation.org/doi/pdf/10.1063/1.1761412}
}

@ARTICLE{Kempski2023_b_field_reversals,
       author = {{Kempski}, Philipp and {Fielding}, Drummond B. and {Quataert}, Eliot and {Galishnikova}, Alisa K. and {Kunz}, Matthew W. and {Philippov}, Alexander A. and {Ripperda}, Bart},
        title = "{Cosmic ray transport in large-amplitude turbulence with small-scale field reversals}",
      journal = {arXiv e-prints},
         year = 2023,
        month = apr,
          eid = {arXiv:2304.12335},
        pages = {arXiv:2304.12335},
          doi = {10.48550/arXiv.2304.12335},
archivePrefix = {arXiv},
       eprint = {2304.12335},
 primaryClass = {astro-ph.HE},
       adsurl = {https://ui.adsabs.harvard.edu/abs/2023arXiv230412335K}
}

@ARTICLE{Sampson2023_SCR_diffusion,
       author = {{Sampson}, Matt L. and {Beattie}, James R. and {Krumholz}, Mark R. and {Crocker}, Roland M. and {Federrath}, Christoph and {Seta}, Amit},
        title = "{Turbulent diffusion of streaming cosmic rays in compressible, partially ionized plasma}",
      journal = {The Monthly Notices of The Royal Astronomical Society},
         year = 2023,
        month = feb,
       volume = {519},
       number = {1},
        pages = {1503-1525},
          doi = {10.1093/mnras/stac3207},
archivePrefix = {arXiv},
       eprint = {2205.08174},
 primaryClass = {astro-ph.GA},
       adsurl = {https://ui.adsabs.harvard.edu/abs/2023MNRAS.519.1503S}
}

@ARTICLE{Zhadankin2013_current_sheet_statistics,
       author = {{Zhdankin}, Vladimir and {Uzdensky}, Dmitri A. and {Perez}, Jean C. and {Boldyrev}, Stanislav},
        title = "{Statistical Analysis of Current Sheets in Three-dimensional Magnetohydrodynamic Turbulence}",
      journal = {The Astrophysical Journal},
         year = 2013,
        month = jul,
       volume = {771},
       number = {2},
          eid = {124},
        pages = {124},
          doi = {10.1088/0004-637X/771/2/124},
archivePrefix = {arXiv},
       eprint = {1302.1460},
 primaryClass = {astro-ph.HE},
       adsurl = {https://ui.adsabs.harvard.edu/abs/2013ApJ...771..124Z}
}

@ARTICLE{Hopkins2021_cr_transport_models,
       author = {{Hopkins}, Philip F. and {Chan}, T.~K. and {Squire}, Jonathan and {Quataert}, Eliot and {Ji}, Suoqing and Keres, Dusan and {Faucher-Gigu{\`e}re}, Claude-Andr{\'e}},
        title = "{Effects of different cosmic ray transport models on galaxy formation}",
      journal = {The Monthly Notices of The Royal Astronomical Society},
         year = 2021,
        month = mar,
       volume = {501},
       number = {3},
        pages = {3663-3669},
          doi = {10.1093/mnras/staa3692},
archivePrefix = {arXiv},
       eprint = {2004.02897},
 primaryClass = {astro-ph.GA},
       adsurl = {https://ui.adsabs.harvard.edu/abs/2021MNRAS.501.3663H}
}

@ARTICLE{Biermann1950_battery,
       author = {{Biermann}, L.},
        title = "{{\"U}ber den Ursprung der Magnetfelder auf Sternen und im interstellaren Raum (miteinem Anhang von A. Schl{\"u}ter)}",
      journal = {Zeitschrift Naturforschung Teil A},
         year = 1950,
        month = jan,
       volume = {5},
        pages = {65},
       adsurl = {https://ui.adsabs.harvard.edu/abs/1950ZNatA...5...65B}
}

@ARTICLE{Gent2021_supernova_turbulence_and_dynamo,
       author = {{Gent}, Frederick A. and {Mac Low}, Mordecai-Mark and {K{\"a}pyl{\"a}}, Maarit J. and {Singh}, Nishant K.},
        title = "{Small-scale Dynamo in Supernova-driven Interstellar Turbulence}",
      journal = {The Astrophysical Journal Letters},
         year = 2021,
        month = apr,
       volume = {910},
       number = {2},
          eid = {L15},
        pages = {L15},
          doi = {10.3847/2041-8213/abed59},
archivePrefix = {arXiv},
       eprint = {2010.01833},
 primaryClass = {astro-ph.GA},
       adsurl = {https://ui.adsabs.harvard.edu/abs/2021ApJ...910L..15G}
}
\bibliographystyle{aasjournal}

\end{document}